\RequirePackage{rotating}
\documentclass[useAMS,usenatbib]{mnras}

\usepackage{ulem}
\usepackage{color}
\usepackage{graphics,graphicx}
\usepackage{times} 
\usepackage{amssymb}
\usepackage{amsmath}
\usepackage{natbib}
\usepackage{lscape}
\usepackage{url}
\usepackage{multirow}
\usepackage{ctable}
\usepackage{threeparttable}
\usepackage[labelfont=bf,labelsep=period]{caption}
\newif\ifAMStwofonts
\AMStwofontstrue

\def\xmm{{\it XMM-Newton}}

\def\swift{{\it Swift}}
\def\swiftng{{\it Neil Gehrels Swift Observatory}}
\def\integral{{\it INTEGRAL}}

\def\epicpn{{EPIC-pn}}
\def\epicmos1{{EPIC-MOS1}}
\def\epicmos2{{EPIC-MOS2}}
\def\epicmos{{EPIC-MOS}}

\def\nustar{{\it NuSTAR}}

\def\deg{$^{\circ}$}

\def\kmps{\hbox{$\rm\thinspace km~s^{-1}$}}
\def\pcmsq{\hbox{$\rm\thinspace cm^{-2}$}}
\def\pcmcub{\hbox{$\rm\thinspace cm^{-3}$}}
\def\H0{{km~s$^{-1}$~Mpc$^{-1}$}}

\def\kev{\hbox{\rm keV}}

\def\ctps{\hbox{$\rm\thinspace ct~s^{-1}$}}

\def\ergpcmsqps{\hbox{$\rm\thinspace erg~cm^{-2}~s^{-1}$}}

\def\ergps{\hbox{erg~s$^{-1}$}}
\def\ergcmps{\hbox{\rm erg~cm~s$^{-1}$}}

\def\msun{\hbox{$M_{\odot}$}}

\def\kte{$kT_{\rm{e}}$}

\def\lambdaCDM{$H_0$ = 70\,\H0, $\Omega_{\rm{M}} = 0.3$, $\Omega_{\Lambda} = 0.7$}

\def\caldb{\textsc{caldb}}

\def\xselect{\textsc{xselect}}

\def\addascaspec{\textsc{addascaspec}}

\def\flx2xsp{\textsc{flx2xsp}}
\def\corner{\textsc{corner}}

\def\nustardas{\textsc{nustardas}}
\def\nupipeline{\textsc{nupipeline}}
\def\nuproducts{\textsc{nuproducts}}

\def\sas{\textsc{sas}}
\def\xmmselect{\textsc{xmmselect}}

\def\epchain{\textsc{epchain}}
\def\emchain{\textsc{emchain}}

\def\rmfgen{\textsc{rmfgen}}
\def\arfgen{\textsc{arfgen}}
\def\epiclccorr{\textsc{epiclccorr}}

\def\rgsproc{\textsc{rgsproc}}
\def\rgscombine{\textsc{rgscombine}}

\def\xrtpipeline{\textsc{xrtpipeline}}
\def\xrtmkarf{\textsc{xrtmkarf}}

\def\carmapack{\textsc{carma\_pack}}

\def\chisq{{$\chi^{2}$}}

\def\xspec{\hbox{\small XSPEC}}
\def\xspecv{\hbox{\small XSPEC}\, v12.6.0f}

\def\xstar{\textsc{xstar}}

\def\tbabs{\textsc{tbabs}}

\def\partcov{\textsc{partcov}}

\def\xillver{\textsc{xillver}}
\def\xillvercp{\textsc{xillver\_cp}}

\def\relxill{\textsc{relxill}}

\def\relxilllpd{\textsc{relxilllpd}}

\def\relxilllpioncp{\textsc{relxilllp\_ion\_cp}}

\def\nthcomp{\textsc{nthcomp}}
\def\cutoffpl{\textsc{cutoffpl}}

\def\ovii{\hbox{\rm O\,{\small VII}}}
\def\oviii{\hbox{\rm O\,{\small VIII}}}
\def\nvii{\hbox{\rm N\,{\small VII}}}

\def\fexxv{\hbox{\rm Fe\,{\small XXV}}}
\def\fexxvi{\hbox{\rm Fe\,{\small XXVI}}}

\def\hbeta{H$\beta$}

\def\eg{{\it e.g.}}

\def\ie{{\it i.e.~\/}}

\def\la{\mathrel{\hbox{\rlap{\hbox{\lower4pt\hbox{$\sim$}}}{\raise2pt\hbox{$<$}}}}}
\def\ga{\mathrel{\hbox{\rlap{\hbox{\lower4pt\hbox{$\sim$}}}{\raise2pt\hbox{$>$}}}}}

\def\d25{D$_{25}$}
\def\nh{{$N_{\rm H}$}}

\def\Ha{{H$\alpha$}}

\def\.25{0.25 keV\thinspace}

\def\lbol{$L_{\rm bol}$}
\def\kbol210{\rm $\kappa_{2-10}$}

\def\mbh{\rm $M_{\rm BH}$}

\def\rg{$R_{\rm{G}}$}
\def\rh{$R_{\rm{H}}$}

\def\Rfrac{$R_{\rm{frac}}$}

\def\obsid{\rm{\small OBSID}}

\def\iras{IRAS\,09149--6206}
\def\nwalker{60}
\def\nstep{30,000}
\def\nburn{5,000}
\def\totchain{1,500,000}

\title[Characterising the SMBH in IRAS\,09149--6206]{A Full Characterisation of the
Supermassive Black Hole in IRAS\,09149--6206}

\author[D.\,J. Walton et al.]
{\parbox{7.in}{D.\,J. Walton$^{1}$ \thanks{E-mail: dwalton@ast.cam.ac.uk},
W. N. Alston$^{1}$,
P. Kosec$^{1}$,
A. C. Fabian$^{1}$,
L. C. Gallo$^{2}$,
J. A. Garcia$^{3,4}$, \\[0.05cm]
J. M. Miller$^{5}$,
E. Nardini$^{6,7}$,
M. T. Reynolds$^{5}$,
C. Ricci$^{8,9,10}$,
D. Stern$^{11}$,
T. Dauser$^{4}$, \\[0.05cm]
F. A. Harrison$^{3}$,
C. S. Reynolds$^{1}$
\\[0.25cm]
\footnotesize
$^{1}$ \it{Institute of Astronomy, University of Cambridge, Madingley Road, Cambridge CB3 0HA, UK} \\
$^{2}$ \it{Department of Astronomy and Physics, Saint Mary’s University, 923 Robie Street, Halifax, NS B3H 3C3, Canada} \\
$^{3}$ \it{Cahill Center for Astronomy and Astrophysics, California Institute of Technology, Pasadena, CA 91125, USA} \\
$^{4}$ \it{Dr. Karl Remeis-Observatory and Erlangen Centre for Astroparticle Physics, Sternwartstr.~7, D-96049 Bamberg, Germany} \\
$^{5}$ \it{Department of Astronomy, University of Michigan, 1085 S. University, Ann Arbor, MI 48109, USA} \\ 
$^{6}$ \it{Dipartimento di Fisica e Astronomia, Universit\`a di Firenze, via G. Sansone 1, I-50019 Sesto Fiorentino, Firenze, Italy} \\
$^{7}$ \it{INAF -- Osservatorio Astrofisico di Arcetri, Largo Enrico Fermi 5, I-50125 Firenze, Italy} \\
$^{8}$ \it{N\'ucleo de Astronom\'ia de la Facultad de Ingenier\'ia, Universidad Diego Portales, Av. Ej\'ercito Libertador 441, Santiago, Chile} \\
$^{9}$ \it{Kavli Institute for Astronomy and Astrophysics, Peking University, Beijing 100871, China} \\
$^{10}$ \it{George Mason University, Department of Physics \& Astronomy, MS 3F3, 4400 University Drive, Fairfax, VA 22030, USA} \\
$^{11}$ \it{Jet Propulsion Laboratory, California Institute of Technology, Pasadena, CA 91109, USA} \\
}}
\date{}

\begin{document}
\pagerange{\pageref{firstpage}--\pageref{lastpage}}
\maketitle
\label{firstpage}

\begin{abstract}
We present new broadband X-ray observations of the type-I Seyfert galaxy \iras, taken
in 2018 with \xmm, \nustar\ and \swift. The source is highly complex, showing a classic
`warm' X-ray absorber, additional absorption from highly ionised iron, strong relativistic
reflection from the innermost accretion disc and further reprocessing by more distant
material. By combining X-ray timing and spectroscopy, we have been able to fully
characterise the supermassive black hole in this system, constraining both its mass and
-- for the first time -- its spin. The mass is primarily determined by X-ray timing constraints
on the break frequency seen in the power spectrum, and is found to be
$\log[M_{\rm{BH}}/M_{\odot}] = 8.0 \pm 0.6$ (1$\sigma$ uncertainties). This is in good
agreement with previous estimates based on the \Ha\ and \hbeta\ line widths, and implies
that \iras\ is radiating at close to (but still below) its Eddington luminosity. The spin is
constrained via detailed modelling of the relativistic reflection, and is found to be $a^* =
0.94^{+0.02}_{-0.07}$ (90\% confidence), adding \iras\ to the growing list of radio-quiet
AGN that host rapidly rotating black holes. The outflow velocities of the various absorption
components are all relatively modest ($v_{\rm{out}} \lesssim 0.03c$), implying these are
unlikely to drive significant galaxy-scale AGN feedback.
\end{abstract}


\section{Introduction}

Supermassive black holes (SMBHs; $M_{\rm{BH}} \gtrsim 10^{6}$\,\msun) are now
thought to lie at the centre of every major galaxy. Accretion onto these black holes is the
primary power source for the variety of different classes of active galactic nuclei (AGN) we
now know of (\citealt{LyndenBell69}). Understanding SMBHs and their accretion is of
particular importance as, despite their disparate size scales, the growth and activity of
these black holes is now understood to play a key role in regulating the formation/evolution
of their host galaxies. This potentially occurs via both their radiative output (\eg\
\citealt{Ishibashi15, Ricci17nat}) and the kinetic output associated with the most powerful
winds (\eg\ \citealt{Pounds03, Tombesi10b, Nardini15, Parker17nat}) and jets (\eg\
\citealt{HlavacekLarrondo12, Ishibashi14}) launched by the accretion process, all of which
is often referred to as `feedback' (see \citealt{Fabian12rev} for a review).

\begin{table*}
  \caption{Details of the 2018 X-ray observations of IRAS\,09149-6206.}
\begin{center}
\begin{tabular}{c c c c c c c c}
\hline
\hline
\\[-0.2cm]
Epoch & Mission & OBSID & Start Date & Exposure\tmark[a] & Raw Count Rate\tmark[b] & Total Counts\tmark[b] \\
& & & & [ks] & [\ctps] & [$\times$1000] \\
\\[-0.3cm]
\hline
\hline
\\[-0.2cm]
\multirow{2}{*}{\vspace{-0.1cm}1} & \xmm\ & 0830490101 & 2018-07-25 & 50/70/71 & 3.6/1.1/0.06 & 180/76/4.2 \\
\\[-0.3cm]
& \nustar\ & 60401020002 & 2018-07-24 & 129 & 0.44 & 56
 \\
\\[-0.1cm]
\multirow{2}{*}{\vspace{-0.1cm}2} & \swift\ & 00088803001 & 2018-08-31 & 1 & 0.24 & 0.2 \\
\\[-0.3cm]
& \nustar\ & 90401630002 & 2018-08-31 & 117 & 0.37 & 42 \\
\\[-0.2cm]
\hline
\hline
\end{tabular}
\end{center}
$^{a}$ \xmm\ exposures are listed for the \epicpn/MOS/RGS detectors; all of the EPIC
detectors were operated in Small Window mode. \\
$^{b}$ Count rates and total counts within our extraction regions are given for
the full band relevant to each detector (0.3--10\,keV for \epicpn/\epicmos/XRT,
7--29\,\AA\ for the RGS, and 3--78\,keV for FPMA/B), and are given per unit for the
\epicmos, RGS and FPM detectors.
\label{tab_obs}
\end{table*}

As such, significant effort has been committed to characterising SMBHs, both in terms of
measuring their masses, particularly via reverberation mapping (\eg\ \citealt{Kaspi00,
Peterson04, Bentz09, Alston20iras}; see \citealt{Peterson14rev} for a review), and their
spin parameters ($a^* = Jc/GM$, where $J$ is the angular momentum of the black hole),
primarily measured by modelling the relativistic reflection from the innermost accretion
disc (\eg\ \citealt{Brenneman11, Gallo11, Risaliti13nat, Walton13spin, Walton14}; see
\citealt{Reynolds14rev} for a review). Mass measurements are key for linking SMBHs to
their host galaxy properties (\eg\ \citealt{Ferrarese06, Kormendy13rev}), as well as
determining how their radiative output scales relative to the Eddington limit (a key
indicator of the mode of accretion), and spin measurements provide information about
their growth history (\eg\ growth via chaotic mergers or prolonged accretion;
\citealt{Sesana14, Fiacconi18}).

\iras\ is a nearby ($z$ = 0.0573) radio-quiet Seyfert-I active galaxy (\citealt{Perez89,
Cram92}). Although it is X-ray bright, detected as part of the hard X-ray surveys
undertaken with the BAT and ISGRI instruments (\citealt{BAT9m, 3ISGRI}) on board the
\swiftng\ (hereafter \swift; \citealt{SWIFT}) and the \integral\ observatory
(\citealt{INTEGRAL}), it has received relatively little dedicated observational attention to
date; prior to this work it has only been the target of a short $\sim$16\,ks observation with
\xmm\ (\citealt{XMM}), and a series of snapshot observations with the \swift\ XRT. These
observations imply the presence of a moderately absorbed AGN, with \nh\ $\sim$
$10^{22}$\,\pcmsq\ (when fit with a neutral absorber; \citealt{Malizia07, Winter09,
Vasudevan10}). However, \cite{Ricci17} find that the majority of the low-energy absorption
is partially ionised, rather than neutral $(\log[\xi/(\rm{erg}~\rm{cm}~\rm{s}^{-1})] \sim 1.5$,
$N_{\rm{H}} \sim 6 \times 10^{22})$\,\pcmsq. In addition to this absorption, and based on
the limited data available to date, \cite{Liebmann18} tentatively note the potential
presence of relativistic disc reflection, and in particular a strong relativistic iron line
(although they do not present any more detailed analysis).

Here we present new broadband X-ray observations of \iras\ taken in 2018 with \xmm,
\nustar\ (\citealt{NUSTAR}) and \swift\ in coordination, from which we are able to place
constraints on both the mass and the spin of its central SMBH.

\section{Observations and Data Reduction}
\label{sec_red}

\nustar\ and \xmm\ performed a coordinated observation of \iras\ in July 2018, and
then \nustar\ performed a further exposure in August 2018, accompanied by a short
snapshot with \swift; a summary of these observations is given in Table \ref{tab_obs}.

\subsection{\nustar}

\begin{figure}
\begin{center}
\hspace*{-0.35cm}
\rotatebox{0}{
{\includegraphics[width=235pt]{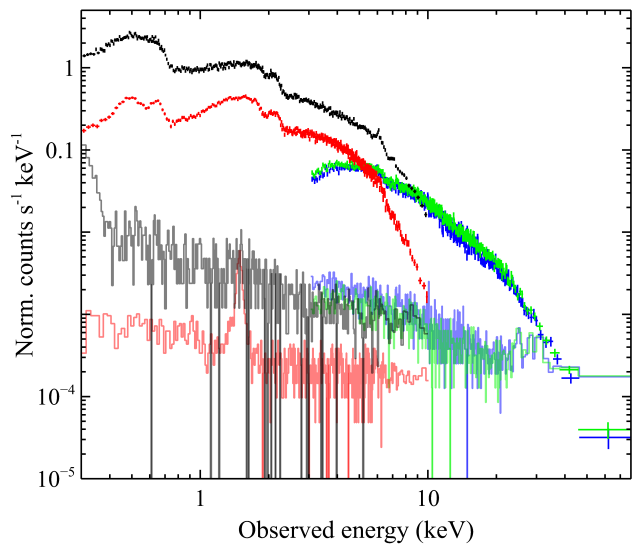}}
}
\end{center}
\vspace*{-0.3cm}
\caption{
The time-averaged \xmm\ and \nustar\ spectra from our coordinated 2018 observation
of \iras\ (epoch 1). The \epicpn\ and \epicmos\ data (\xmm) are shown in black and red,
while the FPMA and FPMB data (\nustar) are shown in green and blue, respectively.
The background levels for each instrument are shown with the solid, stepped lines with
slightly lighter shading than their corresponding source spectra.}
\label{fig_rawspec}
\end{figure}

Each of the two \nustar\ exposures were reduced following standard procedures with
the \nustar\ Data Analysis Software (\nustardas) v1.8.0. For each of the two \nustar\
focal plane modules, FPMA and FPMB, we cleaned the unfiltered event files with
\nupipeline, using instrumental calibration files from the \nustar\ \caldb\ (v20190627).
We used the standard depth correction, which significantly reduces the internal
high-energy background, and excluded passages through the South Atlantic Anomaly
(using the following settings: {\small SAACALC}\,=\,3, {\small SAAMODE}\,=\,Optimized
and {\small TENTACLE}\,=\,yes). Source spectra and lightcurves were extracted from
circular regions of radius 70$''$ using \nuproducts, which was also used to generate
the associated instrumental response files, and background was estimated from
larger regions of blank sky on the same detector as \iras. In order to maximise the
exposure used for spectroscopy, in addition to the standard `science' (mode 1) data,
we also extracted spectra from the `spacecraft science' (mode 6) data following the
procedure outlined in \cite{Walton16cyg}. The mode 6 data provide $\sim$15\% and
$\sim$4\% of the total good exposure for \obsid{s} 60401020002 and 90401630002,
respectively. Although the source flux becomes comparable to the instrumental
background at $\sim$40--50\,keV, the latter is well characterised and \iras\ is detected
across the full \nustar\ band (3--78\,keV; see Figure \ref{fig_rawspec}).

\begin{figure*}
\begin{center}
\hspace*{-0.3cm}
\rotatebox{0}{
{\includegraphics[width=475pt]{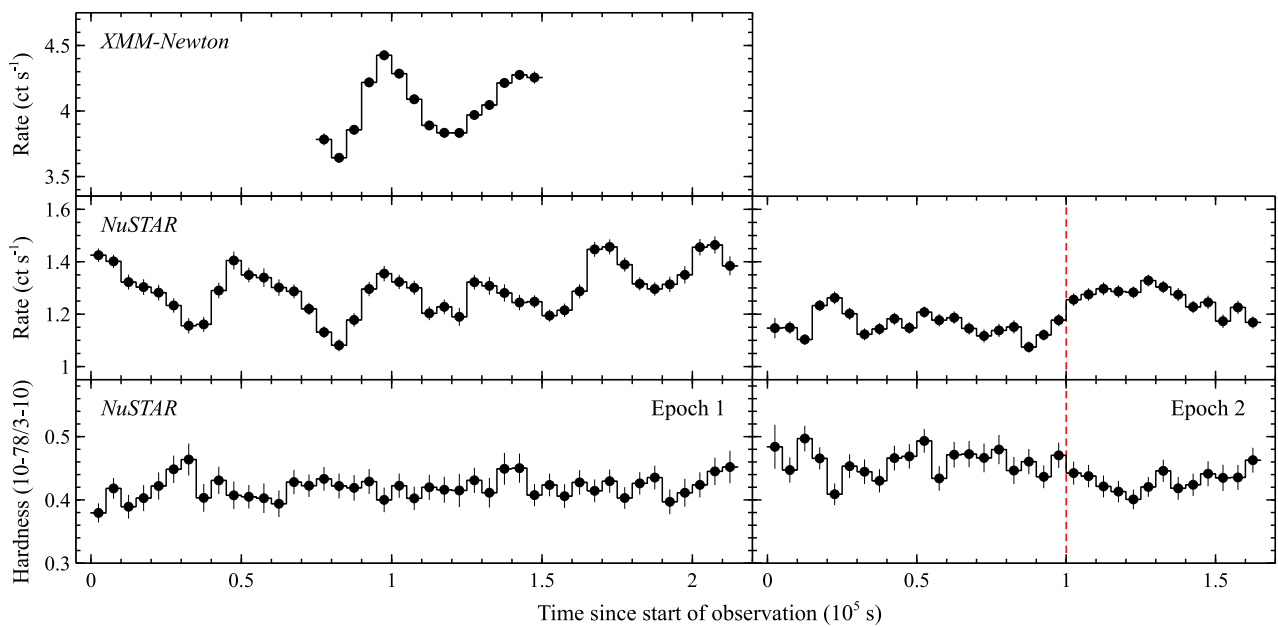}}
}
\end{center}
\vspace*{-0.3cm}
\caption{
The X-ray lightcurves seen by \xmm\ (0.3--10 keV, \epicpn; top panel) and \nustar\
(3--78\,keV, FPMA+FPMB; middle panels) during the 2018 observations of \iras\ (5\,ks
bins). We also show the ratio of the 10--78\,keV and 3--10\,keV bands for the \nustar\
data (bottom panels). Although there is clear flux variability, there is little evidence for
significant spectral variability during epoch 1, and only moderate spectral variability
during epoch 2. The vertical dashed line on the right indicates the point at which we
split the epoch 2 data into epochs 2a and 2b (see Section \ref{sec_2018}).
}
\label{fig_lc}
\end{figure*}

\subsection{\xmm}

The \xmm\ observation presented here was timed to simultaneously overlap with some
portion of the first of the two \nustar\ observations. The reduction of these data was also
carried out following standard procedures, using the \xmm\ Science Analysis System
(\sas\ v18.0.0). 

For the EPIC detectors, we cleaned the raw observation files using \epchain\ and
\emchain\ for the \epicpn\ detector (\citealt{XMM_PN}) and the two \epicmos\ units
(\citealt{XMM_MOS}), respectively. All of the EPIC detectors were operated in Small
Window mode. Source spectra and lightcurves were extracted from the cleaned
eventfiles with \xmmselect\ using a circular region of radius 35$''$. For the \epicpn\
detector, the background was estimated from a larger region of blank sky on the same
detector chip as the source. For the \epicmos\ detectors, the region of the central chip
used in Small Window mode is too small to take a similar approach, so the background
was estimated from large regions of blank sky on adjacent chips. The EPIC data were
free of any significant background flaring, so the whole exposure was used. As
recommended, we only utilized single and double patterned events for \epicpn\
({\small PATTERN}\,$\leq$\,4) and single to quadruple patterned events for \epicmos\
({\small PATTERN}\,$\leq$\,12). The necessary instrumental response files for each of
the detectors were generated using \rmfgen\ and \arfgen, and after performing the
reduction separately for the two \epicmos\ units we also combined these data into a
single spectrum using \addascaspec. Lightcurves are corrected for the PSF losses
using \epiclccorr. The total incident count rates ($\sim$4\,\ctps\ for \epicpn\ and
$\sim$1.4\,\ctps\ for each \epicmos\ unit) were sufficiently low that, given the use of
the Small Window mode, pile-up is of no concern. They are also sufficiently high that
the source flux is always a factor of 10 or more above the background level across the
full EPIC bandpass (again, see Figure \ref{fig_rawspec}).

The data from the Reflection Grating Spectrometer (RGS; \citealt{XMM_RGS}) were
also reduced using \rgsproc, which extracts both the spectral products and their
associated instrumental response files, adopting both the standard source and
background regions. As with the EPIC data, there were no periods of high background
(background rate of $>$ 0.15\,\ctps) in either detector (RGS1/2) and so the full
exposure was used. The net source count rates were $\sim$0.06\,\ctps\ for each RGS
detector, and we merged the data from the two using the \rgscombine\ routine
after confirming there were no notable differences between them over the energies
where both provide coverage.

\subsection{Swift}

For the \swift\ snapshot taken with the second \nustar\ exposure, we extracted the
spectrum from the XRT (\citealt{SWIFT_XRT}). Cleaned event files were generated
with \xrtpipeline\ using the standard filtering, and spectral products were extracted
with \xselect. Source spectra were taken from a circular region of radius $\sim$45'',
and as before the background was estimated from a larger, adjacent region free of
contaminating point sources. The ancilliary response matrix was were generated
with \xrtmkarf, and we use the latest redistribution matrix available in the \swift\
calibration database.

\begin{figure*}
\begin{center}
\hspace*{-0.35cm}
\rotatebox{0}{
{\includegraphics[width=235pt]{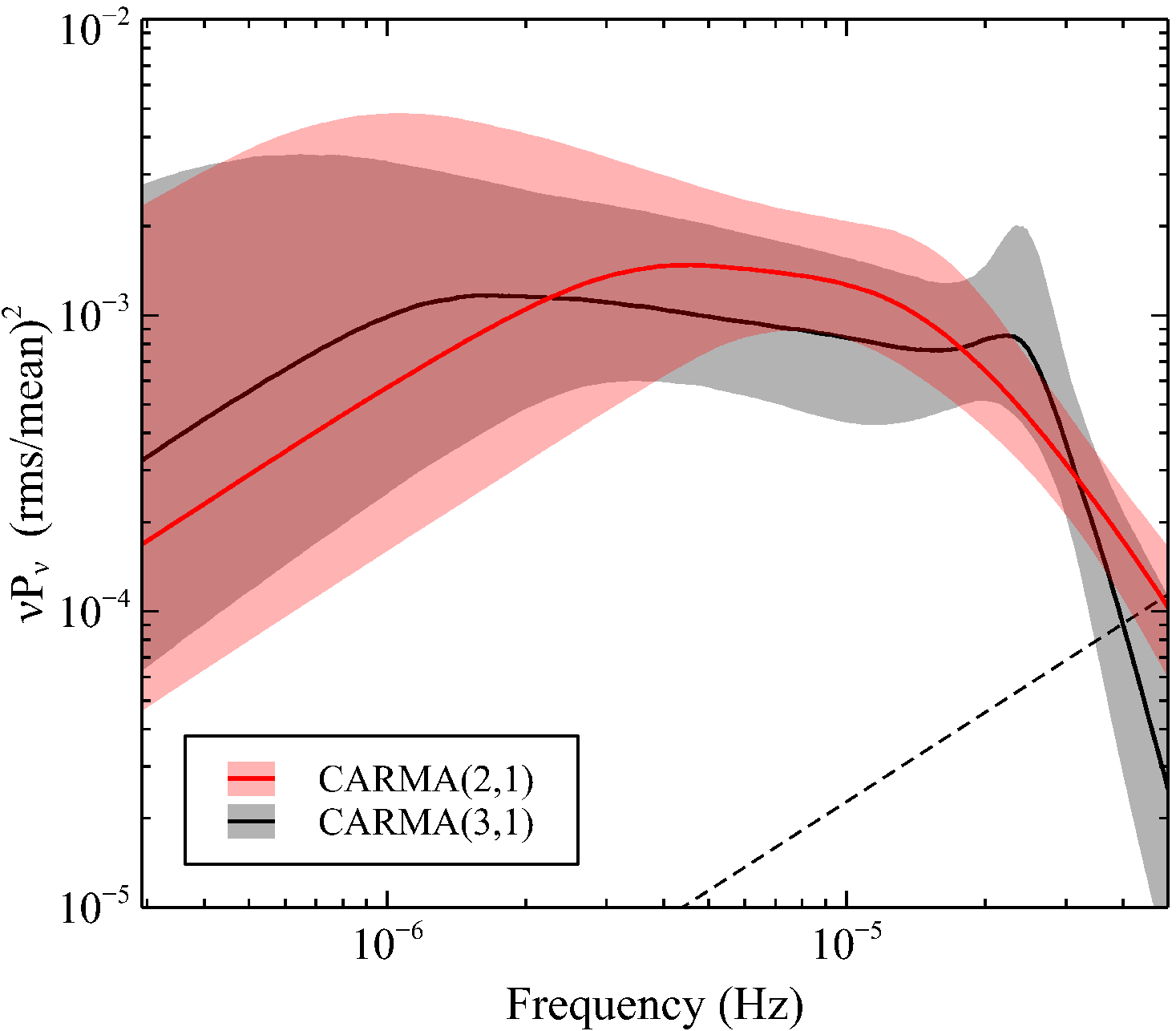}}
}
\hspace{0.5cm}
\rotatebox{0}{
{\includegraphics[width=235pt]{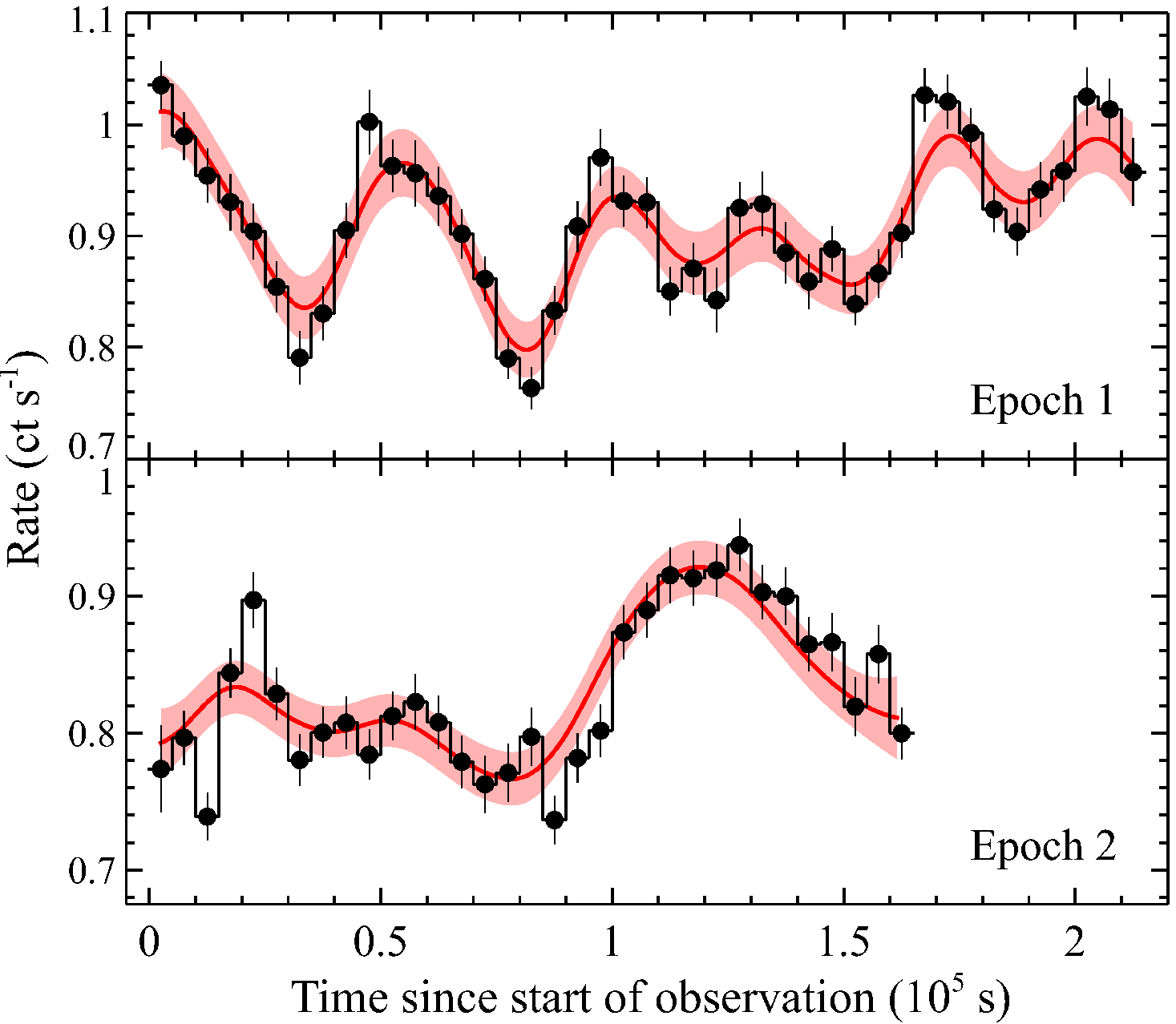}}
}
\end{center}
\vspace*{-0.3cm}
\caption{
\textit{Left panel:} the best-fit CARMA(2,1) and CARMA(3,1) models for the
3--10\,keV PSD of \iras\ (red and black, respectively), based on the \nustar\
lightcurves. The solid lines show the best-fit PSD model in each case, the shaded
regions indicate their $\pm$ 1$\sigma$ uncertainties, based on the uncertainties
on the Lorentzian component parameters from the MCMC chains, and the dashed
line indicates the level of the Poisson noise. \textit{Right panel:} the fit to the 3--10\,keV
\nustar\ lightcurve provided by the CARMA(2,1) model. The black points show the data
(5\,ks time bins, as in Figure \ref{fig_lc}), and the solid red line and shaded area again
show the best-fit model and its $\pm$ 1$\sigma$ uncertainties.}
\label{fig_psd}
\end{figure*}

\section{Analysis}

\subsection{Variability}
\label{sec_var}

We show the \xmm\ and \nustar\ lightcurves from the 2018 observations in Figure
\ref{fig_lc}. Flux variability is clearly seen from \iras\ during the observations presented
here. In particular, one feature that catches the eye in the \nustar\ data from epoch 1 is
a potential quasi-periodic oscillation (QPO) on a timescale of $\sim$40\,ks. As such, we
were granted the second \nustar\ exposure (epoch 2) to see if this behaviour continued,
but there there is no visible indication for the same variations in these data. In order to
investigate whether there are any spectral variations we also compute the hardness ratio
between the 3--10 and 10--78\,keV bands with the \nustar\ data. We see no significant
evidence for spectral changes associated with the flux variability across epoch\,1.
However, during epoch 2 there is some mild variation in the hardness ratio with the
source flux, with the first part of the observation (before an elapsed time of
$\sim$10$^{5}$\,s) slightly fainter and slightly harder than the second part, which is
broadly similar to the epoch\,1 data.

In order to further characterise the variability seen from \iras, particularly in light of the
variations seen in epoch 1, we estimate the power spectral density (PSD) from the \nustar\
data. The sampling of the observations means there are orbital gaps related to the
low-earth orbit of \nustar\ (note that these are not obvious in Figure \ref{fig_lc} owing to the
binning used) as well as a larger gap between the two pointed observations. We therefore
estimated the PSD using the continuous-time autoregressive moving average (CARMA)
method (\citealt{CARMA}) with the public code
\carmapack.\footnote{\url{https://github.com/brandonckelly/carma_pack}} This assumes
the light curve results from a Gaussian noise process and estimates the model power
spectrum as the sum of multiple Lorentzian components, and is well suited to dealing with
non-continuous datasets as it fits the model to the lightcurve data in the time domain.
The two \nustar\ observations are modelled together in order to include the largest number
of cycles for the timescale of interest (\ie $\sim$40\,ks) and to give the best constraints on
the PSD at low frequencies. We considered CARMA($p$,$q$) models, where $p$ is the
number of autoregressive coefficients and $q$ is the number of moving average
coefficients, for a stationary process with $q < p$ (see \citealt{CARMA} for more details).
The Bayesian posterior summaries for the Lorentzian function parameters are formed
using a MCMC sampler. A binsize $dt = 3000$\,s was used giving a total number of bins
$N_{\rm bins} = 126$, but we stress that the results obtained do not depend on the precise
binning used.

As discussed by \cite{Moreno19}, CARMA models with $q \geq 1$ are appropriate for
accreting systems. We therefore consider the two simplest models, CARMA(2,1) and
CARMA(3,1) for the variability exhibited by \iras, and show the resulting power spectra in
Figure \ref{fig_psd}. These have a fairly typical shape for AGN: a slope of $\sim f^{-\alpha}$
with $\alpha > 2$ at frequencies above a characteristic break, $\nu_{\rm b}$, below which
a slope of $\alpha \sim 1$ is observed (\eg\ \citealt{Uttley02, Markowitz03, Papadakis10,
GonzalezMartin12, Alston19iras}). The best-fit CARMA(2,1) and CARMA(3,1) models found
here describe the PSD with a series of either two or three Lorentzians, respectively, while
most prior work modelling AGN PSDs has described them with the broken powerlaw model
described above. Here we assume that the centroid of the highest-frequency Lorentzian in
our PSD model corresponds to the break frequency, as this is the component that
contributes the power around the breaks in Figure \ref{fig_psd} (left). 

The parameters of these Lorentzians are given in Table \ref{tab_psd}. Uncertainties on the
timing parameters are quoted at the 68.3\% level (\ie 1$\sigma$). The extra Lorentzian in
the CARMA(3,1) model is at higher frequencies again than the highest-frequency
component in the CARMA(2,1) model, and the best-fit parameters of this component are
actually fairly narrow and could be considered QPO-like, with the centroid frequency
corresponding to a timescale of $\sim$40\,ks. This component is therefore likely
driven by the variations seen in epoch 1 (as noted previously, these variations are not seen
in epoch 2, although even in the rare cases where AGN QPOs have been robustly detected,
they appear to be transient; \citealt{Alston14qpo}). However, the parameters of this
component are very poorly constrained, and based on the CARMA likelihood fits to the
lightcurves, the addition of this third Lorentzian component is not particularly significant (the
log-likelihoods are 278.0 and 280.1 for the (2,1) and (3,1) models, respectively, giving a
probability of chance improvement for the more complex (3,1) model of $\sim$0.25 based
on a likelihood ratio test with 3 extra free parameters). This is also the case if we consider
epoch 1 by itself. Further observations will be required to determine whether the
CARMA(3,1) model is genuinely a better description of the variability in \iras, and if so to
robustly determine whether the highest frequency component is QPO-like or not. Given this,
we therefore consider the CARMA(2,1) model as our preferred solution at the current time,
and adopt a break frequency of $\nu_{b} = (1.11 \pm 0.55) \times 10^{-5}$\,Hz,
corresponding to a break timescale of $T_{b} = 1.0 \pm 0.5$ days. The fit to the 3--10\,keV
\nustar\ lightcurve provided by this model is shown in Figure \ref{fig_psd} (right).

\begin{table}
  \caption{Parameters for the highest-frequency Lorentzians in the best-fit CARMA(2,1)
  and CARMA(3,1) models for the PSD of \iras. Uncertainties on the timing parameters
  are quoted at the 68.3\% level.}
\begin{center}
\begin{tabular}{c c c c}
\hline
\hline
\\[-0.2cm]
PSD Model & Centroid & Width \\
\\[-0.3cm]
& [$10^{-5}$ Hz] & [$10^{-5}$ Hz] \\
\\[-0.3cm]
\hline
\hline
\\[-0.2cm]
CARMA(2,1) & $1.11 \pm 0.55$ & $2.52^{+2.48}_{-1.64}$ \\
\\[-0.3cm]
CARMA(3,1) & $2.34 \pm 1.35$ & $0.23^{+0.23}_{-0.17}$ \\
\\[-0.2cm]
\hline
\hline
\end{tabular}
\end{center}
\label{tab_psd}
\end{table}

\begin{figure*}
\begin{center}
\hspace*{-0.4cm}
\rotatebox{0}{
{\includegraphics[width=235pt]{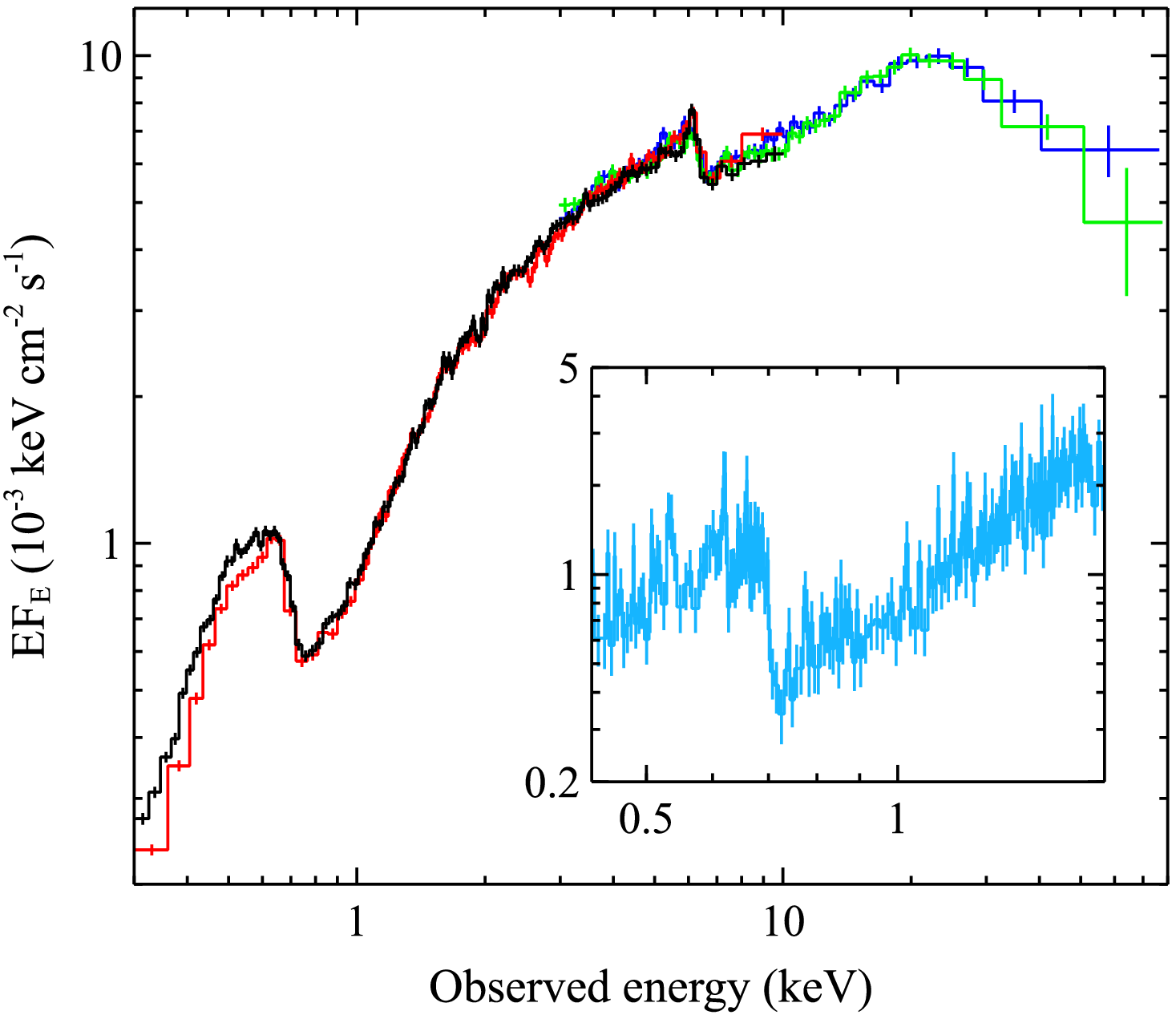}}
}
\hspace*{0.6cm}
\rotatebox{0}{
{\includegraphics[width=235pt]{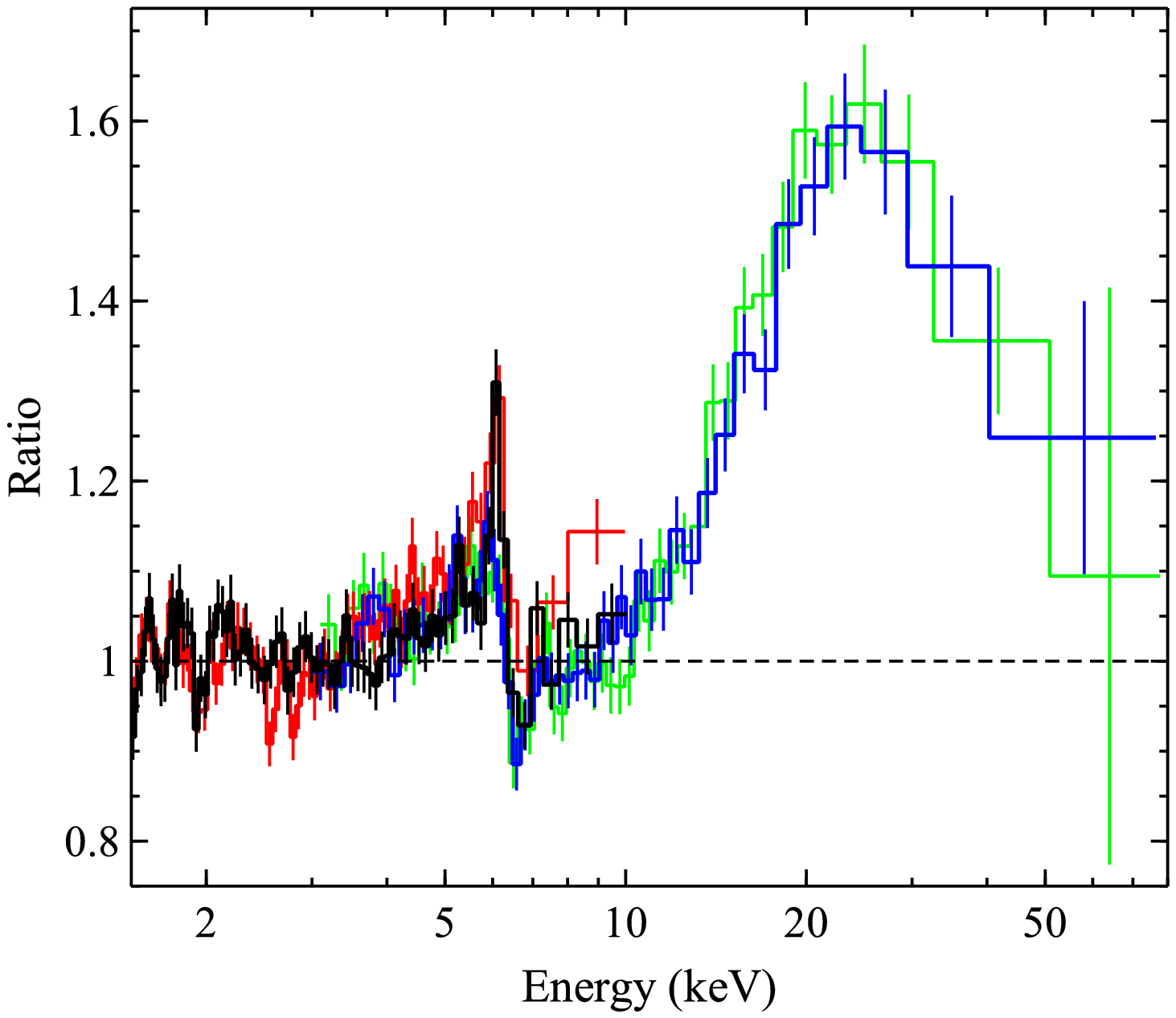}}
}
\end{center}
\vspace*{-0.3cm}
\caption{
\textit{Left panel:} the broadband \xmm+\nustar\ spectrum from epoch 1, after being
unfolded through a model that is constant with energy. Colours in the main panel have
the same meaning as Figure \ref{fig_rawspec}. The data show the source to be
moderately absorbed, with absorption from ionised oxygen in a warm absorber clearly
seen at $\sim$0.7\,keV. The inset shows the \xmm\ RGS data in light blue, confirming
the oxygen absorption. \textit{Right panel:} residuals to a simple \cutoffpl\ continuum,
modified by a partially covering neutral absorber, and applied to the broadband data
over the 1.5--4, 7--10 and 50--78\,\kev\ energy ranges. The key signatures of relativistic
disc reflection are seen: a relativistically broadened iron line at $\sim$6\,\kev\ and a
strong Compton hump at $\sim$20\,keV. The data in all panels have been rebinned for
visual purposes.}
\label{fig_spec}
\end{figure*}

\subsection{Spectroscopy}
\label{sec_spec}

We now present a spectral analysis of the 2018 observations. We use \xspecv\
(\citealt{xspec}) to model the data, and quote uncertainties on the spectral parameters
at the 90\% confidence level for a single parameter of interest. The broadband datasets
(\epicpn\ and \epicmos\ for \xmm, FPMA and FPMB for \nustar) are all binned to a
minimum signal to noise (S/N) of 5 per energy bin, and we fit using \chisq\ minimisation.
The \xmm\ RGS data are binned to a lower level of S/N $\geq$ 3 per bin, in order to
preserve more of the spectral resolution while still being sufficient for \chisq\ minimisation
(note that these S/N requirements are imposed after background subtraction).
Given the relatively limited S/N of the RGS data, we focus on modelling this
simultaneously with the EPIC and \nustar\ datasets in this work. We fit the EPIC data
over the 0.3--10.0\,keV band, the RGS data over the 0.43--1.77\,keV band (7--29\,\AA),
and the \nustar\ data over the 3--78\,keV band. Throughout our analysis we allow
multiplicative constants to vary between the various detectors for data from the same
epoch, primarily in order to account for cross-calibration uncertainties between them.
In the case of the coordinated \xmm+\nustar\ observation, these constants also account
for differences in the average flux level that result from the source variability (Figure
\ref{fig_lc}) and the different temporal coverage of the two exopsures. We fix FPMA at
unity, and the others are found to be within $\sim$15\% of this value. This is similar to
the level of the cross-calibration differences expected between \xmm\ and \nustar\ (flux
differences of $\sim$10\%; \citealt{NUSTARcal}), suggesting that the average flux was
broadly similar across the two exposures, despite their different durations. We initially
focus our spectral analysis on the coordinated \xmm+\nustar\ observation (\ie epoch 1;
Section \ref{sec_XN1}) before proceeding to consider the full 2018 dataset (epochs 1
and 2; Section \ref{sec_2018}).

\subsubsection{The Coordinated XMM-Newton+NuSTAR Observation}
\label{sec_XN1}

Given the lack of variability seen in the hardness ratio during epoch 1 (see Figure
\ref{fig_lc}), we fit these data as a single, time-averaged spectrum. This broadband
spectrum is shown in Figure \ref{fig_spec} (left panel). The source is clearly moderately
absorbed, with a strong oxygen edge from \ovii\ seen at $\sim$0.7\,keV, implying that
the absorbing material is partially ionised, as also concluded by \cite{Ricci17} based on
a previous short \xmm\ observation (and thus is not associated with the interstellar
medium). Such `warm' absorbers are not uncommon in the
X-ray spectra of AGN (potentially seen in $>$50\% of Seyfert galaxies; \eg\
\citealt{Reynolds97wa, Blustin05, Laha14}).

To highlight the features at higher energies, we also show the data/model ratio of the
combined \xmm+\nustar\ data above 1.5\,keV to a simple model consisting of a
\cutoffpl\ continuum with neutral, partially covering absorption (assumed to be at the
redshift of \iras) fit to the 1.5--4, 7--10 and 50--78\,keV bands (here energies are given
in the observed frame) where the primary AGN continuum would be expected to
dominate (Figure \ref{fig_spec}, right panel). For the neutral absorber, we use \tbabs\
(\citealt{tbabs}), adopting the cross-sections of \cite{Verner96} and the solar abundance
set of \cite{Grevesse98} for self-consistency with the \xillver\ reflection models
(\citealt{xillver}) and the \xstar\ photoionisation code (\citealt{xstar}), which are used in
our final, more detailed model for \iras\ (see below). Although the absorption is partially
ionised in reality, this is only supposed to be an illustrative fit, and allowing the neutral
absorber to be partially covering gives it the flexibility to account for the absorption
curvature in the spectrum above $\sim$1.5\,keV. We find a column density of
$N_{\rm{H}} \sim 2.5 \times 10^{22}$\,\pcmsq, a covering factor of $C_{\rm{f}} \sim
0.7$, a photon index of $\Gamma \sim 1.9$ and a cutoff energy of $E_{\rm{cut}} \sim
60$\,keV. This simple model leaves strong residuals in the high-energy portion of the
spectrum. Most notably, a broad emission feature is clearly seen in the iron bandpass,
and a strong excess of emission is also seen above 10\,keV. This high-energy excess
peaks at $\sim$20--30\,keV, as expected for a Compton reflection continuum. As well
as these broad features, a narrower core to the iron emission at $\sim$6\,keV is clearly
visible (corresponding to $\sim$6.4\,keV in the rest-frame of \iras), and evidence for a
narrow absorption feature, most likely from \fexxv, can also be seen at $\sim$6.6\,keV
($\sim$7\,keV rest-frame). Such absorption is also not uncommon in other AGN (\eg\
\citealt{Risaliti05b, Walton18}). However, in addition to these astrophysical features, we
also see evidence for residual features associated with the instrumental edges in the
\xmm\ data at $\sim$2\,keV (in both \epicpn\ and \epicmos). We therefore subsequently
exclude the 1.7--2.5\,keV energy range for these detectors for the rest of our spectral
analysis.

We construct a spectral model in which the intrinsic emission from the central AGN --
which consists of the primary Comptonised X-ray continuum and the associated
relativistic reflection from the inner accretion disc -- is absorbed by a multi-component
warm absorber. We also include a neutral reflector to account for the narrow core of
the iron emission, which is not subject to the warm absorber, and neutral absorption
associated with our own Galaxy, which acts on all emission components. The Galactic
column density towards \iras\ is $N_{\rm{H,Gal}} = 1.58 \times 10^{21}$\,\pcmsq\
(\citealt{NH2016}).

Both the relativistic reflection and the primary continuum from the illuminating X-ray
source are accounted for with the \relxill\ family of models (v1.3.3; \citealt{relxill}). In
particular, we use the \relxilllpioncp\ variant, which self-consistently treats the radial 
emissivity of the disc assuming a lamppost geometry (characterised by the height of
the X-ray source, $h$) and assumes that the primary X-ray continuum is a thermal
Comptonisation spectrum as Compton up-scattering of disc photons is generally
expected to be the physical origin of this emission (\eg\ \citealt{Haardt91}); specifically
the model assumes an \nthcomp\ continuum, characterised by the photon index,
$\Gamma$, and the electron temperature, \kte\ (\citealt{nthcomp1, nthcomp2}).
Although the lamppost model assumes a specific, and simplistic geometry, it is
nevertheless a useful framework as it permits a physical interpretation for the
reflection fraction, \Rfrac\ (see \citealt{relxill_norm} for the definition of \Rfrac\ used in
the latest \relxill\ models), and also allows non-physical regions of parameter space
(\eg\ a very steep radial emissivity profile and a non-rotating black hole) to be
excluded. Following recent work (\citealt{Svoboda12, Kammoun19, Ingram19}), we
also allow for the possibility of an ionisation gradient across the disc, assuming this
has a powerlaw form with radius (characterised by the index $p$ such that $\xi(r)
\propto r^{-p}$), as this allows us to make an agnostic assessment of whether these
effects are important here. We also assume that the inner accretion disc reaches the
innermost stable circular orbit (ISCO) in all our analysis, and fix the outer disk to the
maximum value allowed by the model (1000\,\rg, where \rg = $GM_{\rm{BH}}/c^2$ is
the gravitational radius), and initially we allow \Rfrac\ to vary as a free parameter. The
other key free parameters are the inclination of the disc, its innermost ionisation
parameter, and the iron abundance of the infalling material ($i$, $\xi_{\rm{in}}$ and
$A_{\rm{Fe}}$, respectively; the rest of the elements included in the \xillver/\relxill\
models are assumed to have solar abundances). The ionisation parameter is defined
as standard: $\xi = L_{\rm{ion}}/nR^{2}$, where $L_{\rm{ion}}$ is the ionising
luminosity (integrated over the 0.1--1000\,keV bandpass in \relxill/\xillver), $n$ is the
density of the material, and $R$ is the distance to the ionising source.

The distant reflection is modelled with \xillvercp, as this also assumes an \nthcomp\
input continuum and shares most of its key parameters with \relxilllpioncp. We assume
that the distant reflector is nearly neutral ($\log[\xi/(\rm{erg}~\rm{cm}~\rm{s}^{-1})]$,
the lowest value accepted by \xillvercp) and sees the same ionising continuum as the
disc, after accounting for the gravitational redshift implied by $a^*$ and $h$ in the
lamppost geometry (similar to \citealt{Walton19ufo}). Although \xillvercp\ assumes a
slab geometry, which may not be appropriate for the distant reflector, \cite{Walton18}
found that similar results were obtained for the disc reflection regardless of the
geometry assumed for this emission even in the more absorbed case of
IRAS\,13197-1627. 

Lastly, we use the \xstar\ photoionisation code (\citealt{xstar}) to generate suitable grids
of absorption models for the ionised absorption. We generate two different grids, with the
first designed to model the lower ionisation gas that contributes the oxygen absorption,
and the second designed to model the higher ionisation gas that contributes the iron
absorption. Both grids allow for the ionisation parameter, column density, outflow velocity
and iron and oxygen abundances as free parameters. Note that for \xstar, the bandpass
for the ionising luminosity is defined to be 1--1000\,Ry (\ie 13.6\,eV -- 13.6\,keV). All other
elements have solar abundances. We assume a velocity broadening of 100\,\kmps\ for
the lower ionisation gas (a value typically assumed for such absorption; \eg\
\citealt{Laha14, Longinotti19}), and a velocity broadening of 3000\,\kmps\ for the higher
ionisation gas (also motivated by the broadening used in previous work on similar
absorbers; \eg\ \citealt{Risaliti05b, Walton18}). We assume a fairly generic ionising
continuum of $\Gamma = 2$ in both cases to allow for broader applicability; this is
reasonably close to the typical X-ray spectrum for unobscured AGN (\citealt{Ricci17}).
For self-consistency, we link the iron abundance parameters across all the different
model components associated with \iras. We also link the oxygen abundances for all of
the ionised absorption components (this is not currently a free parameter in the
\xillver/\relxill\ models).

\begin{table}
  \caption{Results obtained for the lamppost reflection model fit to the broadband
  \xmm+\nustar\ data for \iras. Uncertainties on the spectral parameters are quoted at the
  90\% level.}
\begin{center}
\begin{tabular}{c c c c}
\hline
\hline
\\[-0.2cm]
Component & \multicolumn{3}{c}{Parameter} \\
\\[-0.25cm]
\hline
\hline
\\[-0.15cm]
{\small WA1} & $\log\xi$ & $\log$[\ergcmps] & $1.12^{+0.04}_{-0.06}$ \\
\\[-0.3cm]
& $N_{\rm{H}}$ & [$10^{22}$ cm$^{-2}$] & $1.00^{+0.07}_{-0.04}$ \\
\\[-0.3cm]
& $A_{\rm{O}}$ & [solar] & $1.24^{+0.08}_{-0.06}$ \\
\\[-0.3cm]
& $v_{\rm{out}}$ & [\kmps] & $4200 \pm 400$ \\
\\[-0.3cm]
& $C_{\rm{f}}$ & [\%] & $81^{+2}_{-3}$  \\
\\
{\small WA2} & $\log\xi$ & $\log$[\ergcmps] & $2.00^{+0.01}_{-0.02}$ \\
\\[-0.3cm]
& $N_{\rm{H}}$ & [$10^{22}$ cm$^{-2}$] & $6.2^{+0.2}_{-0.3}$ \\
\\[-0.3cm]
& $v_{\rm{out}}$ & [\kmps] & $7300^{+400}_{-500} $ \\
\\[-0.3cm]
& $C_{\rm{f}}$ & [\%] & $67^{+1}_{-2}$ \\
\\
{\small HIA} & $\log\xi$ & $\log$[\ergcmps] & $3.44^{+0.04}_{-0.06}$ \\
\\[-0.3cm]
& $N_{\rm{H}}$ & [$10^{22}$ cm$^{-2}$] & $6.5^{+1.2}_{-1.4}$ \\
\\[-0.3cm]
& $v_{\rm{out}}$ & [\kmps] & $9300 \pm 1000$ \\
\\
\relxill\tmark[a] & $\Gamma$ & & $2.16 \pm 0.02$ \\ 
\\[-0.3cm]
& $kT_{\rm{e}}$\tmark[b] & [keV] & $90^{+80}_{-30}$ \\
\\[-0.3cm]
& $a^*$ & & $0.94^{+0.02}_{-0.06}$ \\
\\[-0.3cm]
& $i$ & [\deg] & $42^{+2}_{-1}$ \\
\\[-0.3cm]
& $h$ & [\rg] & $3.6^{+1.2}_{-0.5}$ \\
\\[-0.3cm]
& \Rfrac\ & & $2.1 \pm 0.2$ \\
\\[-0.3cm]
& $\log\xi_{\rm{in}}$ & $\log$[\ergcmps] & $1.9 \pm 0.2$ \\
\\[-0.3cm]
& $p$ & & $0.10^{+0.23}_{-0.05}$ \\
\\[-0.3cm]
& $A_{\rm{Fe}}$ & [solar] & $1.8 \pm 0.1$ \\
\\[-0.3cm]
& Norm & [$10^{-4}$] & $3.9^{+1.0}_{-0.3}$ \\
\\
\xillver\tmark[a] & Norm & [$10^{-5}$] & $2.0^{+0.4}_{-0.5}$ \\
\\[-0.15cm]
\hline
\\[-0.2cm]
\chisq/DoF & & & 3336/3201 \\
\\[-0.25cm]
\hline
\hline
\end{tabular}
\label{tab_param_XN1}
\end{center}
\flushleft
$^a$ We use the \relxilllpioncp\ and \xillvercp\ variants here. \\
$^b$ $kT_{\rm{e}}$ is quoted in the rest-frame of the illuminating X-ray source (\ie prior
to any gravitational redshift), based on the best-fit lamppost geometry. \\
\vspace{0.4cm}
\end{table}

\begin{figure}
\begin{center}
\hspace*{-0.35cm}
\rotatebox{0}{
{\includegraphics[width=235pt]{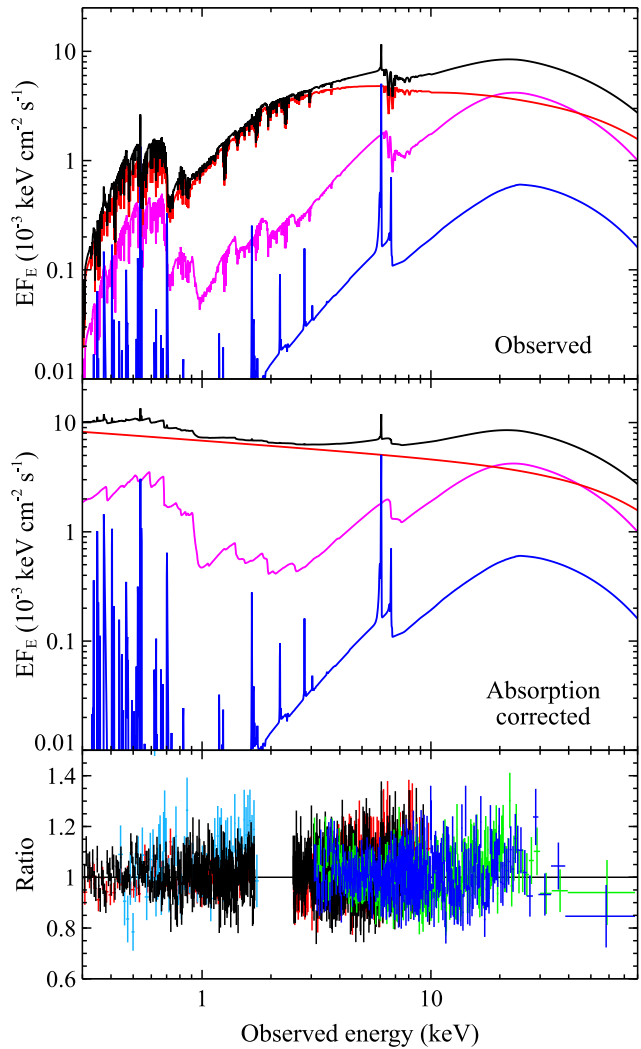}}
}
\end{center}
\vspace*{-0.3cm}
\caption{\textit{Top panel:} the relative contributions of the different components for
our broadband spectral model for the coordinated \xmm+\nustar\ observation of \iras\
(epoch 1). The total model is shown in black, the Comptonised continuum in red, the
relativistic disc reflection in magenta, and the distant reflection in blue. \textit{Middle
panel:} same as the top panel, but with all of the absorption components removed.
\textit{Bottom panel:} The data/model ratio for our broadband fit. The data have been
rebinned for visual purposes, and the colours have the same meanings as in Figure
\ref{fig_spec}.}
\label{fig_model}
\end{figure}

\begin{figure*}
\begin{center}
\hspace*{-0.4cm}
\rotatebox{0}{
{\includegraphics[width=235pt]{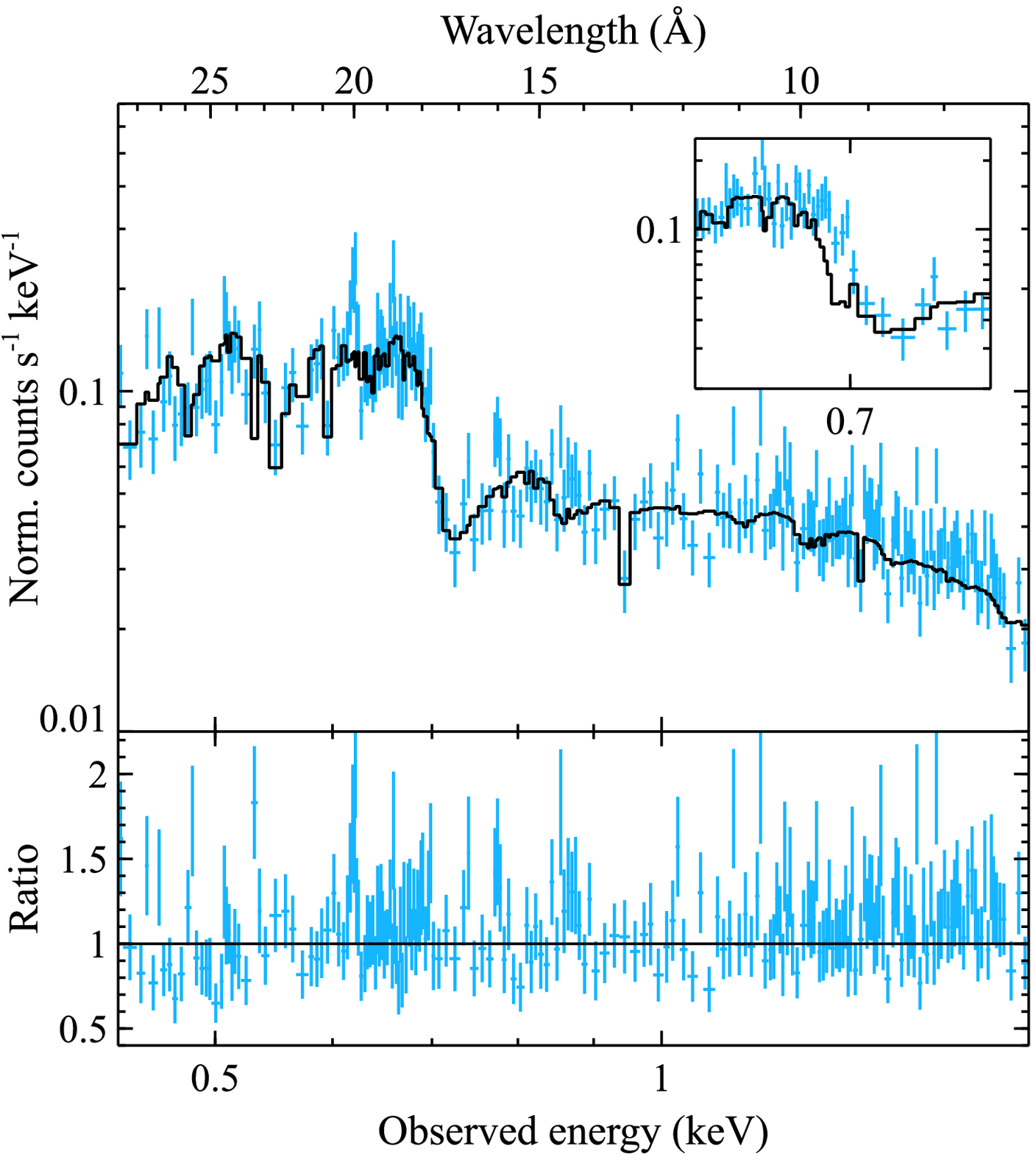}}
}
\hspace{0.6cm}
\rotatebox{0}{
{\includegraphics[width=235pt]{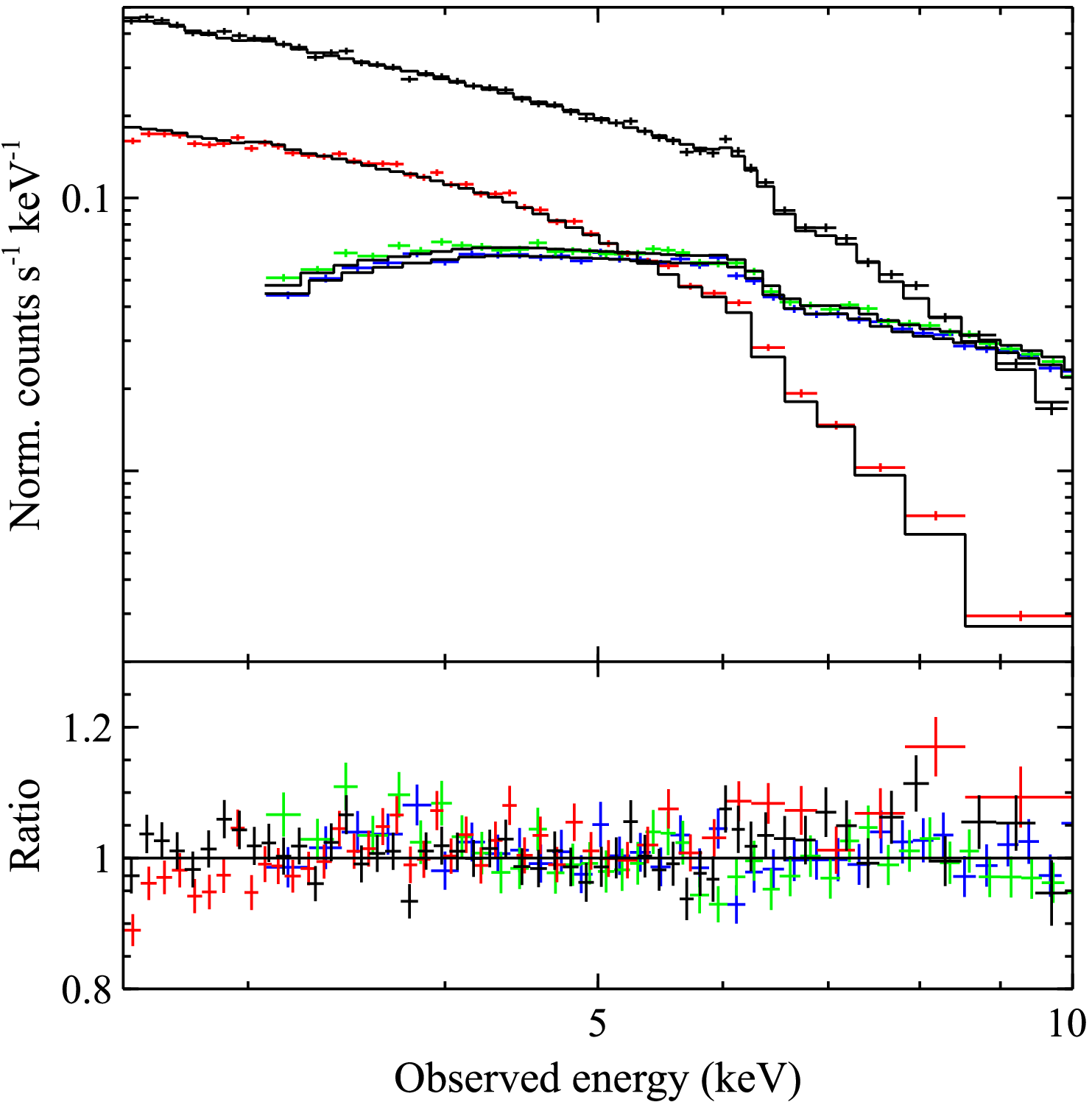}}
}
\end{center}
\vspace*{-0.3cm}
\caption{Zoomed in fits to the \xmm\ RGS data (\textit{left panels}) and the iron K
bandpass (\textit{right panels}). For each of these energy ranges we show the fits in
count space (\textit{top panels}), and the corresponding data/model ratios
(\textit{bottom panels}). In all cases, the data are shown in the same colours as Figure
\ref{fig_spec}, with the same binning, and the total model is shown with the stepped
black line. The inset in the top-left panel shows the result of setting the outflow
velocities of the WA components to zero (while holding all other parameters constant);
in this case the model clearly misses the position of the main oxygen absorption edge.
}
\label{fig_fit}
\end{figure*}

During our analysis, we allow the lower ionisation \xstar\ absorption to be partially
covering using the \partcov\ model within \xspec\ (the \xstar\ grids themselves are not
calculated to include $C_{\rm{f}}$ as a free parameter, and assume this to be unity). We
also find that the low-energy oxygen absorption is best described with a combination of
two \xstar\ components with different ionisation parameters, the first ({\small WA1})
contributes the majority of the \ovii\ absorption (0.73\,keV rest-frame), and the second
({\small WA2}) contributes most of the \oviii\ absorption (0.87\,keV rest-frame). This is
more complex than the absorption model used previously by \cite{Ricci17}, but we
stress that the S/N of the \xmm\ data used in that work is significantly lower than the S/N
of the data presented here. The higher ionisation absorption ({\small HIA}) is instead
assumed to be fully covering for simplicity; this component essentially only contributes
the iron absorption line at $\sim$6.6\,keV (observed frame), so the covering factor and
the column density are fully degenerate if both are allowed to vary. We note that with
this treatment of the ionised absorption, we do not find the need for any further neutral
component associated with \iras. Our final model expression is as follows:
\tbabs$_{\rm{Gal}} \times$ $($\xillvercp\ + {\small WA1} $\times$ {\small WA2} $\times$
{\small HIA} $\times$ \relxilllpioncp$)$, where we note again that {\small WA1} and
{\small WA2} are both partially covering. We stress that the removal of any of these
components significantly degrades the fit (by $\Delta\chi^{2} \gtrsim 20$ per degree of
freedom). Although we have assumed that the ionised absorption components
do not apply to the distant reflection, we also note that making the alternative
assumption (\ie that they do) does not significantly change the quality of the fits, or
result in any changes in the key model parameters of interest. We have also investigated
allowing for different values of $\Gamma$ for the \xmm\ and \nustar\ data (\eg\
\citealt{Cappi16, Middei18}), which could potentially result from subtly different
calibrations for the two missions. However, we find that this does not make a large
difference to the fit ($\Delta\chi^{2} = 14$ for one more free parameter) and does not
introduce significant changes in any of the key parameters of interest, so we present the
model with $\Gamma$ linked between \xmm\ and \nustar.

\begin{figure}
\begin{center}
\hspace*{-0.35cm}
\rotatebox{0}{
{\includegraphics[width=235pt]{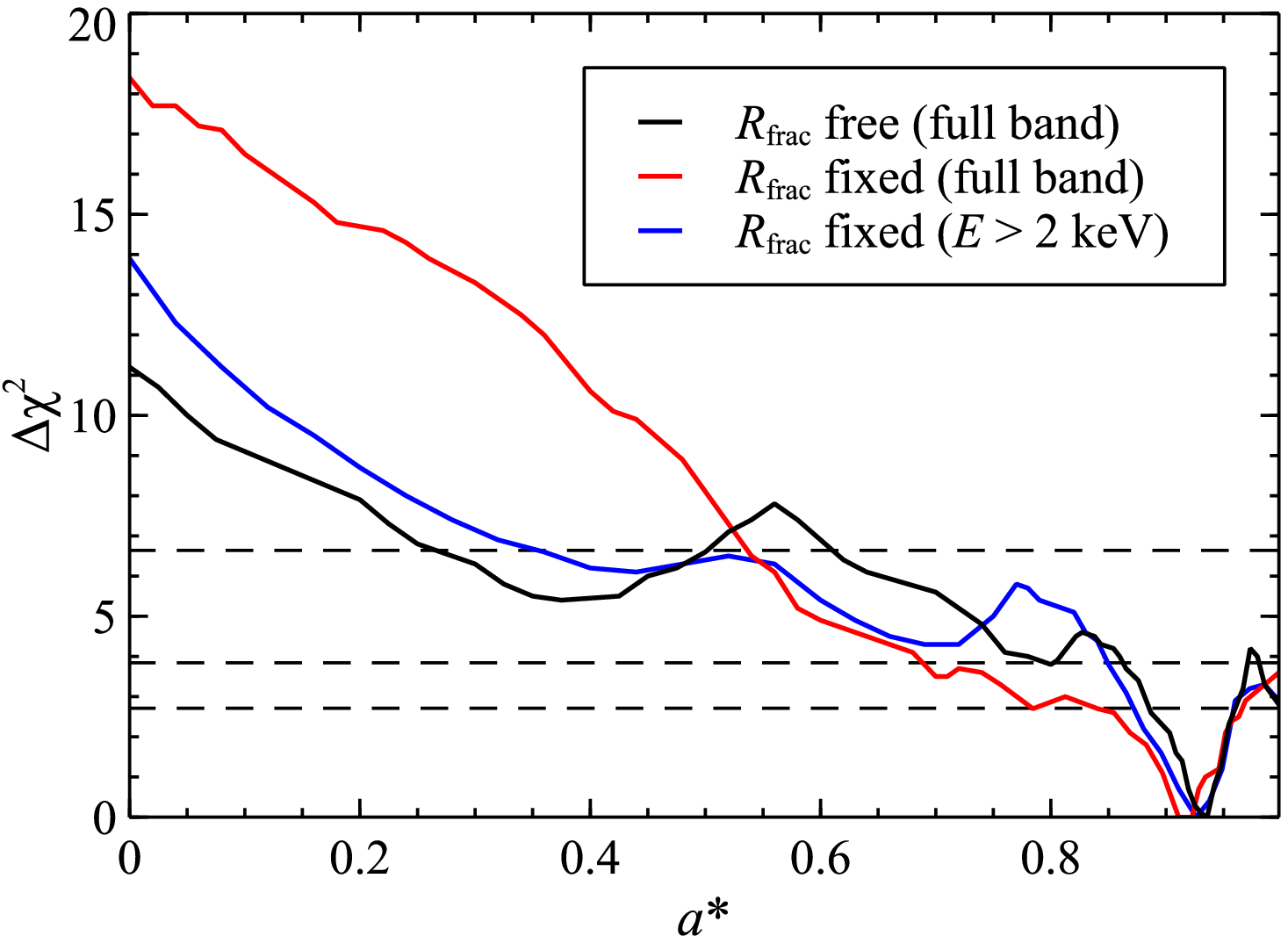}}
}
\end{center}
\vspace*{-0.3cm}
\caption{
The $\Delta$\chisq\ confidence contours for the spin of \iras\ based on our spectral
modeling of the coordinated \xmm+\nustar\ observation (epoch 1). We show contours
for our models with \Rfrac\ free to vary (black) and computed self-consistently from
$a^*$ and $h$ in the lamppost geometry (red). The horizontal dotted lines represent
the 90, 95 and 99\% confidence levels for a single parameter of interest. We also show
the contour for the latter case based on just the data above 2\,keV (blue).}
\label{fig_spin_XN1}
\end{figure}

\begin{figure*}
\begin{center}
\hspace*{-0.35cm}
\rotatebox{0}{
{\includegraphics[width=250pt]{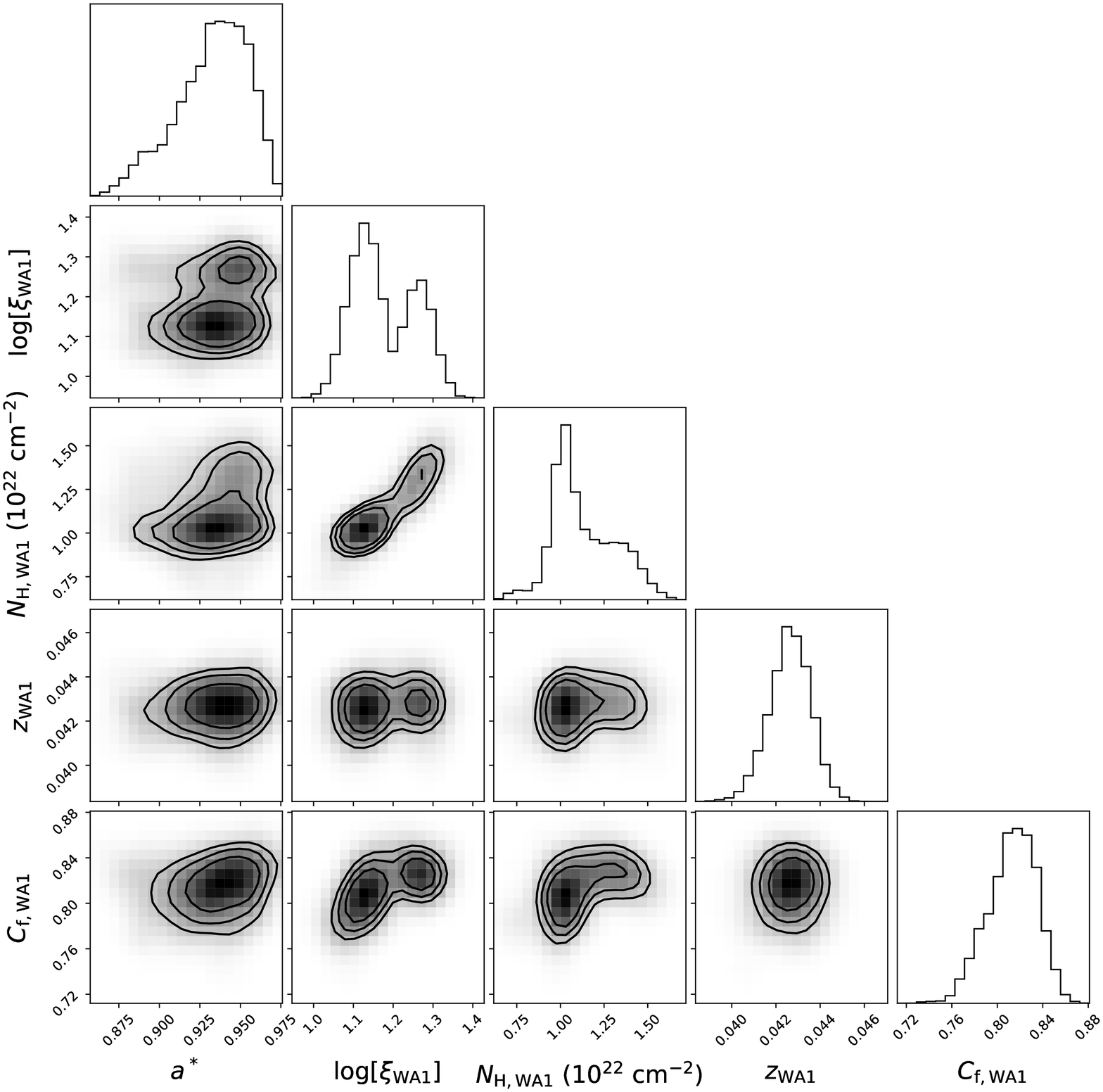}}
}
\rotatebox{0}{
{\includegraphics[width=250pt]{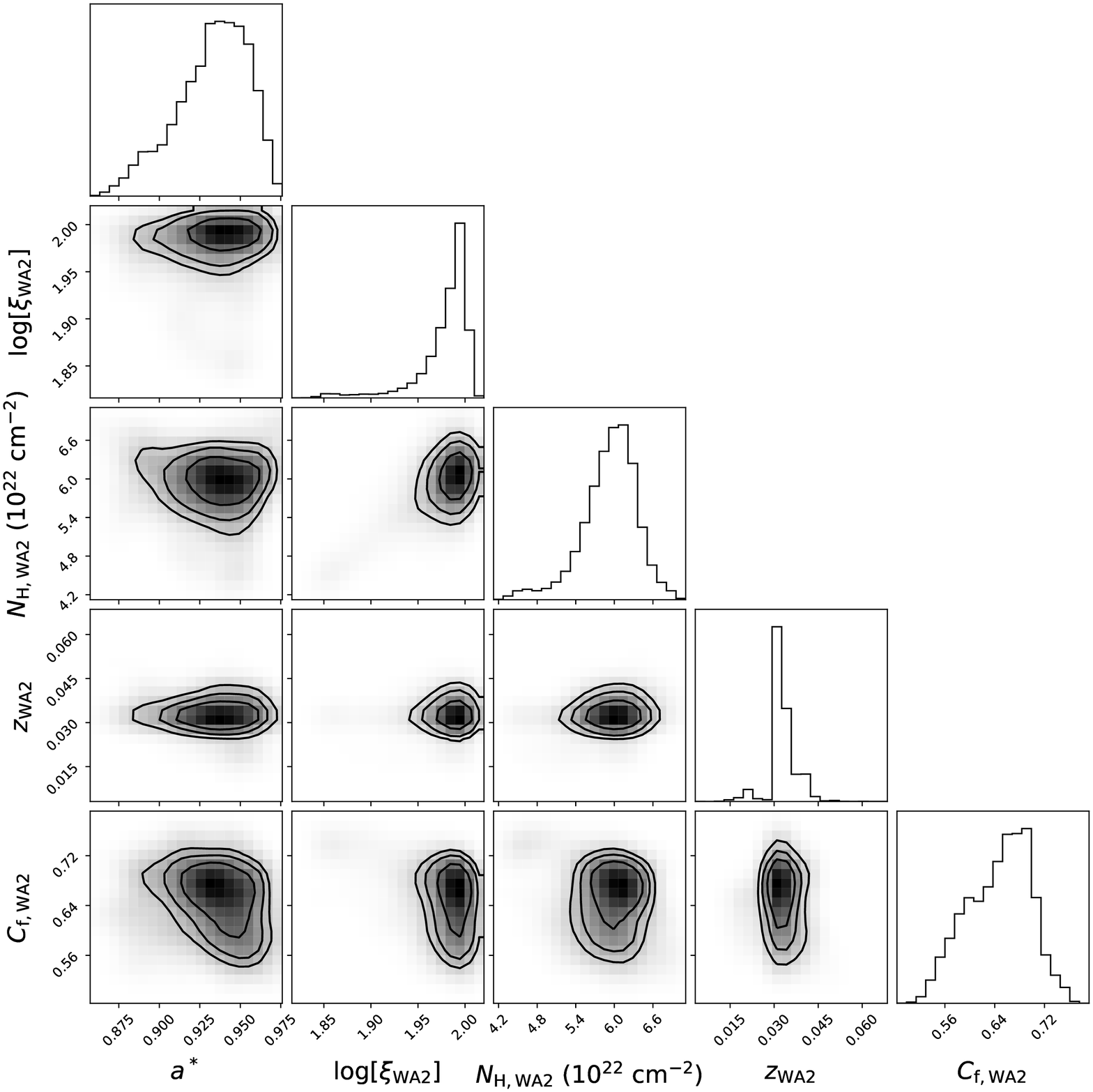}}
}
\end{center}
\vspace*{-0.3cm}
\caption{
The MCMC results for the black hole spin vs the parameters for the two main warm
absorber components (\textit{left} -- WA1; \textit{right} -- WA2) included in our model
for epoch 1. Note that  the outflow velocities of the absorbers are given here in terms
of their redshifts in the observed frame. The 2-D contours show the 1, 2 and 3$\sigma$
confidence levels.}
\label{fig_mcmc}
\end{figure*}

\begin{figure}
\begin{center}
\hspace*{-0.35cm}
\rotatebox{0}{
{\includegraphics[width=235pt]{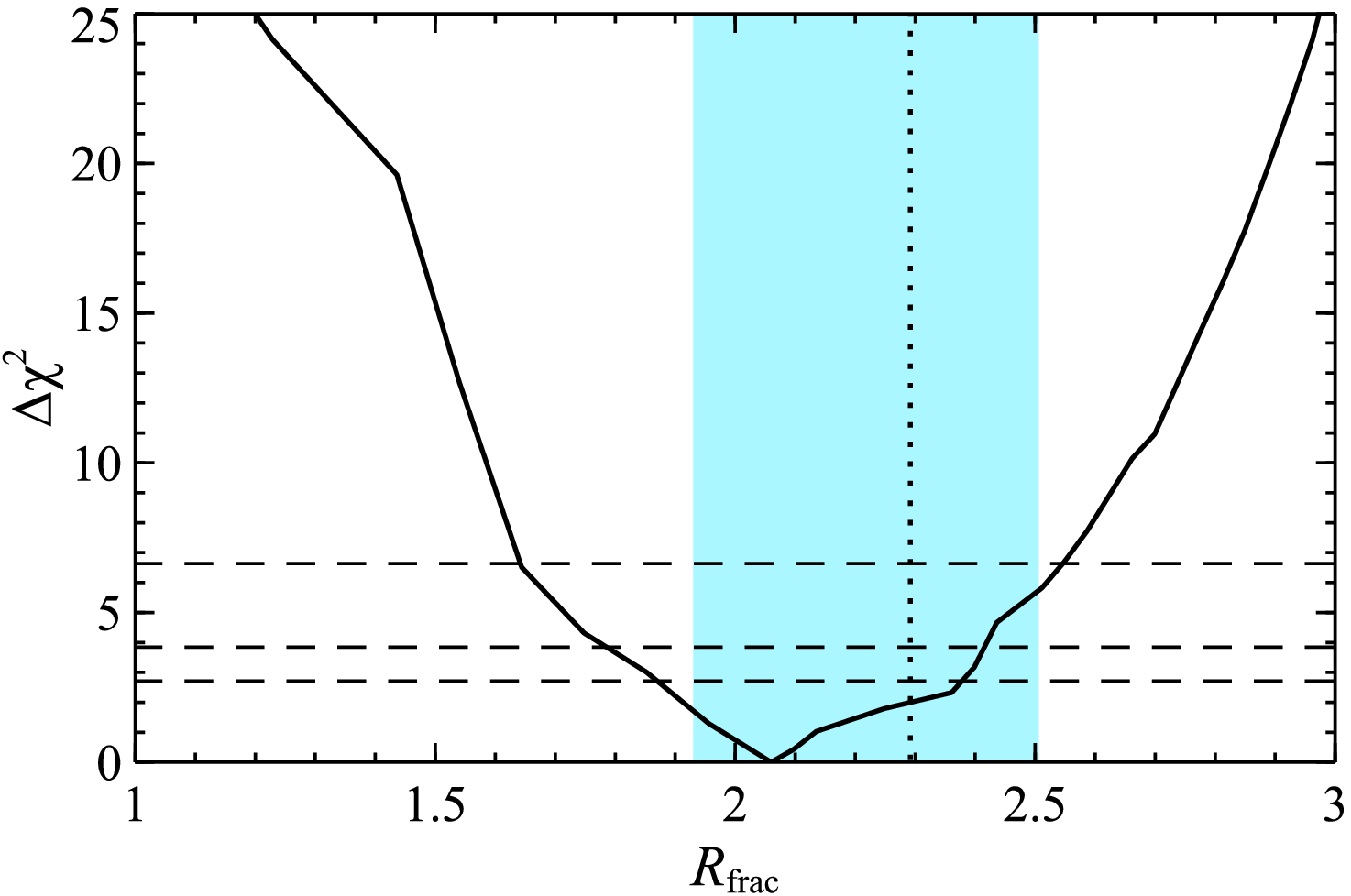}}
}
\end{center}
\vspace*{-0.3cm}
\caption{
The $\Delta$\chisq\ confidence contours for \Rfrac\ when varied as a free parameter
in our analysis of epoch 1 (solid black). The horizontal dotted lines represent the same
confidence levels as Figure \ref{fig_spin_XN1}. The vertical dotted line indicates the
predicted value of \Rfrac\ based on $a^*$ and $h$ in the lamppost geometry, and the
shaded region indicates the range predicted by the 90\% statistical uncertainties on
these parameters.}
\label{fig_Rfrac}
\end{figure}

This model describes the \iras\ data from epoch 1 well, with \chisq\ = 3336 for 3201
degrees of freedom (DoF), and the best-fit parameters are given in Table
\ref{tab_param_XN1}. The relative contributions of the various model components --
both with and without the line-of-sight absorption -- are shown in Figure \ref{fig_model},
along with the corresponding broadband data/model ratio, showing that the model
reproduces the broadband spectral shape well. We also show zoomed in fits for the
\xmm\ RGS data and the iron K bandpass in Figure \ref{fig_fit}, demonstrating the
quality of fit in these key areas of the spectrum. We find that even when allowing for
complex, partially covering, partially ionised absorption, the data still require a strong
contribution from relativistic reflection from the innermost accretion disc. In particular,
we find that the spin of the black hole is high, $a^* = 0.94^{+0.02}_{-0.06}$, and the
X-ray source is compact, $h = 3.6^{+1.2}_{-0.5}$\,\rg; we show the constraints on the
spin in Figure \ref{fig_spin_XN1}. One potential concern when fitting complex spectral
models similar to that utilized here relates to degeneracies between different model
parameters. In addition to our standard \chisq\ analysis, we therefore also perform a
series of Monte Carlo Markoff Chain (MCMC) simulations to provide a further
exploration of the best-fit parameter space. In particular, we make use of the MCMC
functionality within \xspec, and explore the parameter space using the
Goodman-Weare algorithm (\citealt{MCMC_GW}) and the best-fit model as a starting
point. All model parameters reported in Table \ref{tab_param_XN1} are free to vary
throughout this analysis. We use \nwalker\ walkers, each run for \nstep\ steps with a
burn-in length of \nburn, resulting in a total chain of \totchain\ parameter combinations.
Chain convergence is good, with the convergence measure proposed by
\cite{Geweke92} close to zero for every parameter. Here we focus on investigating
whether there are any strong dependences between the spin parameter and the
ionised absorbers in our model, since these play a major role in sculpting the
observed broadband spectrum; further parameter combinations are presented in
Appendix \ref{app_mcmc}. We find that there are no strong degeneracies between the
spin and the properties of the ionised absorption components; for illustration we plot
the 2-D parameter constraints from our MCMC simulations for the spin vs the key
parameters for the two main warm absorber components (WA1, WA2) in Figure
\ref{fig_mcmc}, but we stress that the same conclusion would be drawn for any of the
other absorption parameters. Furthermore, the 90\% uncertainty on the spin implied
by these simulations is $a^* = 0.94^{+0.02}_{-0.05}$, in excellent agreement with our
\chisq\ analysis; we therefore continue with the latter in the further analysis described
below.

The best-fit reflection fraction is quite large, \Rfrac\ = $2.1 \pm 0.2$, as expected for
a rapidly rotating black hole with a compact corona. In fact, the best-fit reflection
fraction actually matches that predicted from the combination of $a^*$ and $h$ in the
lamppost geometry remarkably well (predicted \Rfrac\ = $2.3^{+0.2}_{-0.4}$, based
on the statistical constraints on $a^*$ and $h$; see Figure \ref{fig_Rfrac}). We
therefore re-fit the data computing \Rfrac\ self-consistently from $a^*$ and $h$; we
do not report these fits in detail, since the results for the other key parameters are all
consistent with those presented in Table \ref{tab_param_XN1}, but the updated
constraints on the black hole spin are also shown in Figure \ref{fig_spin_XN1}. The
formal spin constraints are also similar, $a^* = 0.91^{+0.04}_{-0.05}$, but here we
find that low spin values are excluded at a much higher level of confidence. We also
note that although we allow for a radial ionisation gradient, the data do not require
one, as the constraints are consistent with $p = 0$ in both cases; at most they only
allow for a fairly shallow gradient, with $p < 0.34$. This may be due to the compact
nature of the corona inferred, which will in turn result in the reflected emission
primarily arising from the innermost regions of the disc. 

\begin{figure}
\begin{center}
\hspace*{-0.35cm}
\rotatebox{0}{
{\includegraphics[width=235pt]{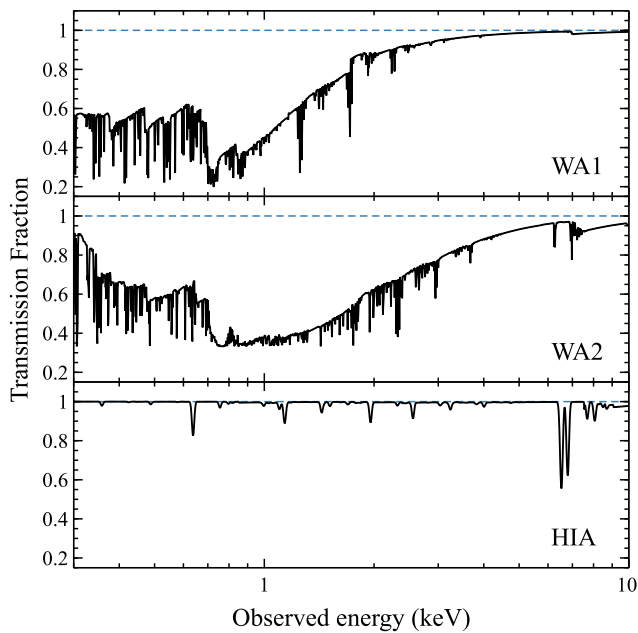}}
}
\end{center}
\vspace*{-0.3cm}
\caption{
Transmission profiles for each of the three individual absorption components included in
our spectral model for \iras\ (where a value of 1 indicates 100\% of the incident emission
is transmitted). All of the parameters for each of the individual absorbers ($\xi$, \nh,
$v_{\rm{out}}$, $C_{\rm{f}}$) are as quoted in Table \ref{tab_param_XN1}. The
combination of the WA1 and WA2 components dominate the oxygen absorption at low
energies; WA1 contributes the majority of the \ovii\ absorption, while WA2 contributes
the majority of the \oviii\ absorption. The HIA component models the ionised iron
absorption as a blend of \fexxv\ and \fexxvi\ absorption lines.
}
\label{fig_abs}
\end{figure}

Although we find evidence that the iron abundance is mildly super-solar, we also note
that the best-fit oxygen abundance for the ionised absorption is close to the solar
value. As such, even though this is not a free parameter for the reflection models, there
are no issues relating to significantly different abundances between the different
components. The column densities and ionisation states of the absorption components
are relatively typical for such warm absorbers; we show the transmission profile for
each of the absorption components in Figure \ref{fig_abs}). It is worth noting that the
best-fit photon index for \iras\ of $\Gamma \sim 2.15$ is slightly steeper than that
assumed when initially calculating the \xstar\ grids. As the definition of the ionisation
parameter in \xstar\ is based on a bandpass that extends to significantly lower energies
than our X-ray data, for steeper ionising continua higher global ionisation parameters
would be required to produce the same number of ionising photons in the X-ray band,
and so our ionisation parameters will be systematically underestimated to some extent.
To quantify this, we also calculate a small \xstar\ grid around the best-fit parameters of
the WA2 component assuming $\Gamma = 2.15$ (and otherwise the same setup as
described above); using this grid for WA2 instead we find that the difference in
ionisation parameter is only $\Delta\log\xi \sim 0.1$. The other model parameters are
all identical to the best-fit values reported in Table \ref{tab_param_XN1}.

The outflow velocities found for WA1 and WA2 are relatively high for such absorption,
but similar velocities have still been reported previously for outflows with similar ionisation
states to those seen here (\eg\ \citealt{Laha14, Longinotti19}), and forcing the WA
components to have no outflow velocity clearly misses the position of the Oxygen edge
(see Figure \ref{fig_fit}). We also see evidence for increasing outflow velocities with
increasing ionisation parameter, potentially suggesting we are looking at radially stratified
absorbers (broadly similar to that seen by \citealt{Kosec18_1h0707} in emission in the
narrow line Seyfert 1 1H\,0707-495, albeit seen in absorption and at more modest
outflow velocities here). This is in part because the data strongly prefer a solution in
which the iron absorption is a blend of \fexxv\ and \fexxvi\ with the \xstar\ grid used here
(see Figure \ref{fig_abs}). We test this potential stratification further by repeating the fits
after linking the outflow velocities of the different absorption components in various
combinations. Forcing the velocity of the WA2 component to be the same as either the
WA1 or HIA components (such that there are now only two distinct velocity components)
only provides a mild degradation of the fit ($\Delta\chi^{2}$ =7--8 for one less free
parameter in both cases), so it is plausible that WA2 could represent a distinct ionisation
phase of either of these other two kinematic outflow components (\eg\ \citealt{Reeves20}).
In both of these scenarios, the key inner disc reflection parameters remain consistent with
those presented in Table \ref{tab_param_XN1}. However, forcing all of the intrinsic
absorption components (WA1, WA2, HIA) to have a common velocity does result in a
significantly worse fit ($\Delta\chi^{2} = 29$ for two fewer free parameters), so the data do
clearly prefer at least some velocity structure to the absorption.

The best-fit absorption model predicts a variety of weak narrow features throughout the
spectrum, in addition to the dominant Oxygen structure. However, these are mostly
either outside of the RGS band, or the current RGS data does not have sufficient S/N to
detect them individually. The only other feature associated with the ionised absorption
clearly seen in the RGS data is the \nvii\ edge (0.67\,keV/18.5\,\AA\ rest-frame) seen at
$\sim$0.63\,keV (the edge at  $\sim$0.55\,keV/22.5\,\AA\ is associated with the Galactic
column). There is also some mild evidence in the RGS data for a narrow emission line at
0.61\,keV, which would correspond to \oviii\ (0.65\,keV/19.1\,\AA rest-frame) at the
redshift of \iras, and therefore re-emission from the WA2 component (which has the
larger column of the two lower-ionisation warm absorbers). We therefore investigate
including a photoionised emitter -- also calculated with \xstar\ in the same way as WA1/2
-- to represent re-emission from WA2 (\ie with the column density, ionisation parameter
linked to those of WA2 and the iron and oxygen abundances linked to the rest of the
model components). However, this only results in a relatively moderate improvement in
the fit statistic, with $\Delta\chi^{2} = 12$ for one more free parameter, and the addition
of this component does not change any of the other key model parameters, so we do
not include this in the final model. 


\begin{figure}
\begin{center}
\hspace*{-0.35cm}
\rotatebox{0}{
{\includegraphics[width=235pt]{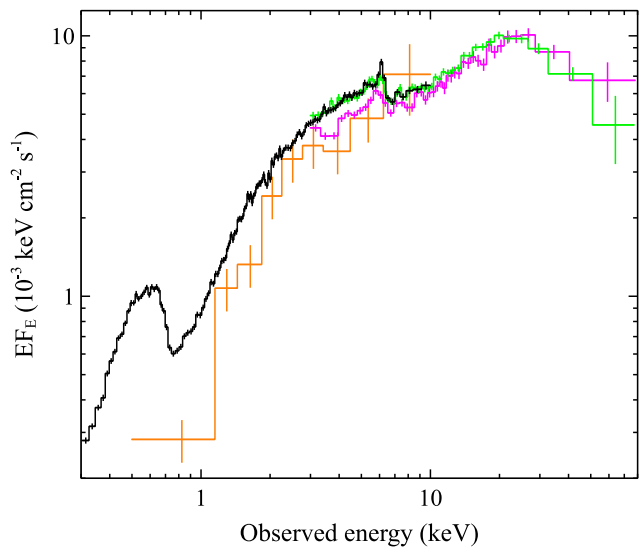}}
}
\end{center}
\vspace*{-0.3cm}
\caption{
A comparison of the spectra of \iras\ from epochs 1 and 2a. For clarity, we only show the
\epicpn, XRT and FPMA datasets, unfolded through a model that is constant with energy
(as in Figure \ref{fig_spec}). For epoch 1, the colours match Figure \ref{fig_spec}, while
for epoch 2a the XRT data are shown in orange, and the FPMA data in magenta. 
During epoch 2a, \iras\ exhibits a slightly harder spectrum than the rest of the data (\ie
epochs 1 and 2b, which show practically identical spectra).}
\label{fig_epochs}
\end{figure}

\subsubsection{The Combined 2018 Dataset}
\label{sec_2018}

Having established our best-fit model for epoch 1, we now perform a combined
fit including the \swift+\nustar\ data from epoch 2. As noted previously, in contrast to epoch
1 there appears to be some mild but systematic spectral variability during epoch 2, with
the first part of the \nustar\ observation slightly harder than the second (see Figure
\ref{fig_lc}), and the second part showing basically identical hardness to epoch 1. We
therefore split the \nustar\ data, extracting separate spectra from the periods before and
after an elapsed time of $T_{\rm{obs}} = 10^5$\,s. Owing to the low-earth orbit of \nustar,
these spectra, which we refer to epochs 2a and 2b, have exposures of $\sim$72 and
44\,ks, respectively. The short 1\,ks \swift\ exposure taken along with the \nustar\
observation occurred during the first part of the observation (epoch 2a), while epoch 2b
has no corresponding soft X-ray coverage. We show a comparison of the broadband
spectrum from epochs 1 and 2a in Figure \ref{fig_epochs}; as indicated from the simple
hardness ratios shown in Figure \ref{fig_lc}, the spectrum from epoch 2a is slightly harder
than epoch 1 (the spectra from epoch 2b are identical to epoch 1, also as indicated by the
hardness ratios, and so are not shown for clarity).

\begin{table}
  \caption{Results obtained for the lamppost reflection model fit to the full 2018 dataset
  for \iras. Uncertainties on the spectral parameters are quoted at the 90\% level.}
\begin{center}
\begin{tabular}{c c c c}
\hline
\hline
\\[-0.2cm]
Component & \multicolumn{3}{c}{Parameter} \\
\\[-0.25cm]
\hline
\hline
\\[-0.1cm]
\multicolumn{4}{c}{\textit{Epoch 1:}} \\
\\[-0.2cm]
{\small WA1} & $\log\xi$ & $\log$[\ergcmps] & $1.12 \pm 0.05$ \\
\\[-0.3cm]
& $N_{\rm{H}}$ & [$10^{22}$ cm$^{-2}$] & $1.00^{+0.09}_{-0.08}$ \\
\\[-0.3cm]
& $A_{\rm{O}}$ & [solar] & $1.25 \pm 0.08$ \\
\\[-0.3cm]
& $v_{\rm{out}}$ & [\kmps] & $4200^{+300}_{-400}$ \\
\\[-0.3cm]
& $C_{\rm{f}}$ & [\%] & $82^{+1}_{-2}$ \\
\\
{\small WA2} & $\log\xi$ & $\log$[\ergcmps] & $2.00^{+0.01}_{-0.02}$ \\
\\[-0.3cm]
& $N_{\rm{H}}$ & [$10^{22}$ cm$^{-2}$] & $6.2^{+0.2}_{-0.4}$ \\
\\[-0.3cm]
& $v_{\rm{out}}$ & [\kmps] & $7300^{+300}_{-400}$ \\
\\[-0.3cm]
& $C_{\rm{f}}$ & [\%] & $66 \pm 3$ \\
\\
{\small HIA} & $\log\xi$ & $\log$[\ergcmps] & $3.46 \pm 0.04$ \\
\\[-0.3cm]
& $N_{\rm{H}}$ & [$10^{22}$ cm$^{-2}$] & $7.7^{+2.7}_{-1.4}$ \\
\\[-0.3cm]
& $v_{\rm{out}}$ & [\kmps] & $9000 \pm 1000$ \\
\\
\relxill\tmark[a] & $\Gamma$ & & $2.16 \pm 0.01$ \\ 
\\[-0.3cm]
& $kT_{\rm{e}}$\tmark[b] & [keV] & $70^{+30}_{-20}$ \\
\\[-0.3cm]
& $a^*$ & & $0.94^{+0.02}_{-0.07}$ \\
\\[-0.3cm]
& $i$ & [\deg] & $43^{+3}_{-2}$ \\
\\[-0.3cm]
& $h$ & [\rg] & $4.0^{+1.0}_{-0.3}$ \\
\\[-0.3cm]
& \Rfrac\tmark[c] & & $2.2^{+0.1}_{-0.3}$ \\
\\[-0.3cm]
& $\log\xi_{\rm{in}}$ & $\log$[\ergcmps] & $1.7^{+0.3}_{-0.4}$ \\
\\[-0.3cm]
& $p$ & & $<0.34$ \\
\\[-0.3cm]
& $A_{\rm{Fe}}$ & [solar] & $1.7^{+0.2}_{-0.1}$ \\
\\[-0.3cm]
& Norm & [$10^{-4}$] & $3.7^{+0.2}_{-0.4}$ \\
\\
\xillver\tmark[a] & Norm & [$10^{-5}$] & $2.0^{+0.4}_{-0.5}$ \\
\\[-0.15cm]
\hline
\\[-0.1cm]
\multicolumn{4}{c}{\textit{Epoch 2a: \tmark[d]}} \\
\\[-0.2cm]
{\small WA1} & $N_{\rm{H}}$ & [$10^{22}$ cm$^{-2}$] & $2.5^{+0.6}_{-1.1}$ \\
\\
{\small WA2} & $N_{\rm{H}}$ & [$10^{22}$ cm$^{-2}$] & $2.0^{+2.4}_{-1.3}$ \\
\\
{\small HIA} & $v_{\rm{out}}$\tmark[e] & [\kmps] & $3000 \pm 2000$ \\
\\
\relxill\tmark[a] & $\Gamma$ & & $2.07 \pm 0.02$ \\
\\[-0.3cm]
& Norm & [$10^{-4}$] & $2.7^{+0.2}_{-0.5}$ \\
\\[-0.15cm]
\hline
\\[-0.1cm]
\multicolumn{4}{c}{\textit{Epoch 2b: \tmark[d]}} \\
\\[-0.2cm]
\relxill\tmark[a] & Norm & [$10^{-4}$] & $3.5^{+0.4}_{-0.5}$ \\
\\[-0.15cm]
\hline
\\[-0.2cm]
\chisq/DoF & & & 4749/4583 \\
\\[-0.25cm]
\hline
\hline
\end{tabular}
\label{tab_param_2018}
\end{center}
\flushleft
$^a$ We use the \relxilllpioncp\ and \xillvercp\ variants here. \\
$^b$ $kT_{\rm{e}}$ is quoted in the rest-frame of the illuminating X-ray source (\ie prior
to any gravitational redshift), based on the best-fit lamppost geometry. \\
$^c$ \Rfrac\ is calculated self-consistently for the lamppost geometry from
$a$ and $h$; the errors represent the range of values permitted by varying these
parameters within their 90\% uncertainties. \\
$^d$ Parameters not listed for epochs 2a and 2b are linked to their corresponding
parameters from epoch 1, unless noted otherwise (see text). \\
$^e$ The outflow velocity of the HIA is linked for epochs 2a and 2b.
\vspace{0.4cm}
\end{table}

\begin{figure}
\begin{center}
\hspace*{-0.35cm}
\rotatebox{0}{
{\includegraphics[width=235pt]{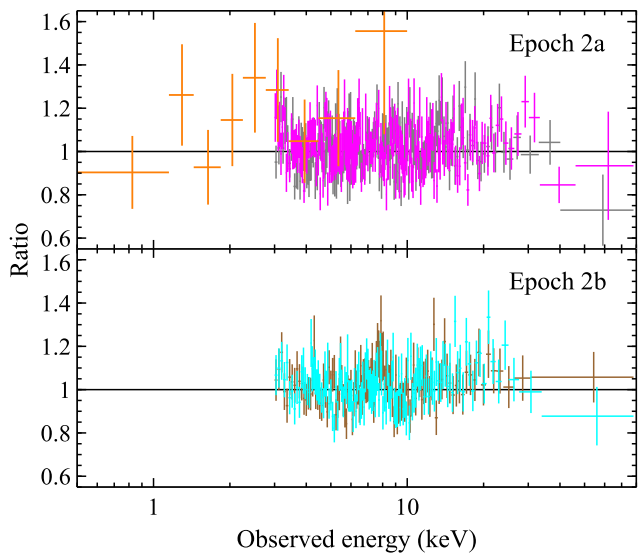}}
}
\end{center}
\vspace*{-0.3cm}
\caption{
The data/model ratio for the data from epochs 2a and 2b with our fit to the full 2018
dataset. For epoch 2a, the XRT, FPMA and FPMB data are shown in orange, magenta
and grey, respectively, matching Figure \ref{fig_epochs} (where relevant), and for
epoch 2b the FPMA and FPMB data are shown in cyan and brown, respectively. The
data have been rebinned for visual purposes.}
\label{fig_ratio_e2}
\end{figure}

We model the full 2018 dataset (epochs 1, 2a and 2b) simultaneously, with the model
constructed in Section \ref{sec_XN1}. For these fits, we retain the self-consistent
treatment of \Rfrac\ in the lamppost geometry, given the results seen for epoch 1. Other
key physical parameters that should not vary on observational timescales are linked
across all datasets: the spin, the inclination, the iron and oxygen abundances, and the
normalisation of the distant reflector. For practical purposes, given either the low S/N
or lack of soft X-ray coverage available for epoch 2, we also link a variety of other
parameters between the different epochs: although there is some flux variability
associated with the spectral variability, this is very mild (the observed 2--10\,keV flux
varies by $\sim$15\%), so we also link all of the various ionisation parameters across
the different epochs. Furthermore, given both the lack of any soft X-ray coverage and
the similarity of the \nustar\ spectra, we link all of the parameters for the warm
absorber components (WA1, WA2) between epochs 1 and 2b.

\begin{table*}
  \caption{Observed fluxes for the full model and absorption-corrected fluxes for the
  \relxill\ component during the 2018 observations of \iras\ considered here for several
  (rest-frame) bandpasses.}
\begin{center}
\begin{tabular}{c c c c c c c c c}
\hline
\hline
\\[-0.2cm]
Epoch & \multicolumn{4}{c}{Observed Fluxes (full model)} & \multicolumn{4}{c}{Absorption-Corrected Fluxes (\relxill)} \\
& \multicolumn{4}{c}{[$10^{-11}$\,\ergpcmsqps]} & \multicolumn{4}{c}{[$10^{-11}$\,\ergpcmsqps]} \\
& 2--10\,keV & 0.3-10.0\,keV & 10--80\,keV & 0.3--80\,keV & 2--10\,keV & 0.3-10.0\,keV & 10--80\,keV & 0.3--80\,keV \\
\\[-0.3cm]
\hline
\hline
\\[-0.2cm]
1 & $1.54 \pm 0.01$ & $1.91 \pm 0.02$ & $2.57 \pm 0.06$ & $4.48^{+0.07}_{-0.06}$ & $1.93 \pm 0.02$ & $5.0 \pm 0.2$ & $2.48^{+0.08}_{-0.06}$ & $7.5 \pm 0.2$ \\
\\[-0.3cm]
2a & $1.32 \pm 0.02$ & $1.57^{+0.07}_{-0.05}$ & $2.58^{+0.08}_{-0.07}$ & $4.1 \pm 0.1$ & $1.57 \pm 0.04$ & $3.7 \pm 0.2$ & $2.42^{+0.09}_{-0.07}$ & $6.1 \pm 0.2$ \\
\\[-0.3cm]
2b & $1.48 \pm 0.02$ & $1.83 \pm 0.02$ & $2.47 \pm 0.06$ & $4.30^{+0.08}_{-0.04}$ & $1.85 \pm 0.03$ & $4.8 \pm 0.2$ & $2.38 \pm 0.07$ & $7.2 \pm 0.2$ \\
\\[-0.2cm]
\hline
\hline
\end{tabular}
\end{center}
\label{tab flux}
\end{table*}

With this initial setup, we then explored which other parameters were consistent with
remaining constant across the different epochs. When this occurred, we linked these
parameters in our final combined fit to the data. The height of the X-ray source, the
gradient of the radial ionisation profile of the accretion disc, the electron temperature
of the primary continuum emission, and the column density of the HIA component were
all found to be consistent with remaining constant across all epochs. The outflow
velocities and the covering factors of both the warm absorber components (WA1, WA2)
were consistent with remaining constant between epochs 1 and 2a (and so are
effectively kept constant for all epochs). The photon indices were found to vary between
epochs 1 and 2a, but were consistent for epochs 1 and 2b. Some evidence for variability
in the column densities of the two warm absorber components (WA1, WA2) between
epochs 1 and 2a is also seen, and the outflow velocity of the HIA component was found
to vary between epochs 1 and 2 (but was consistent across epochs 2a and 2b).

\begin{figure}
\begin{center}
\hspace*{-0.35cm}
\rotatebox{0}{
{\includegraphics[width=235pt]{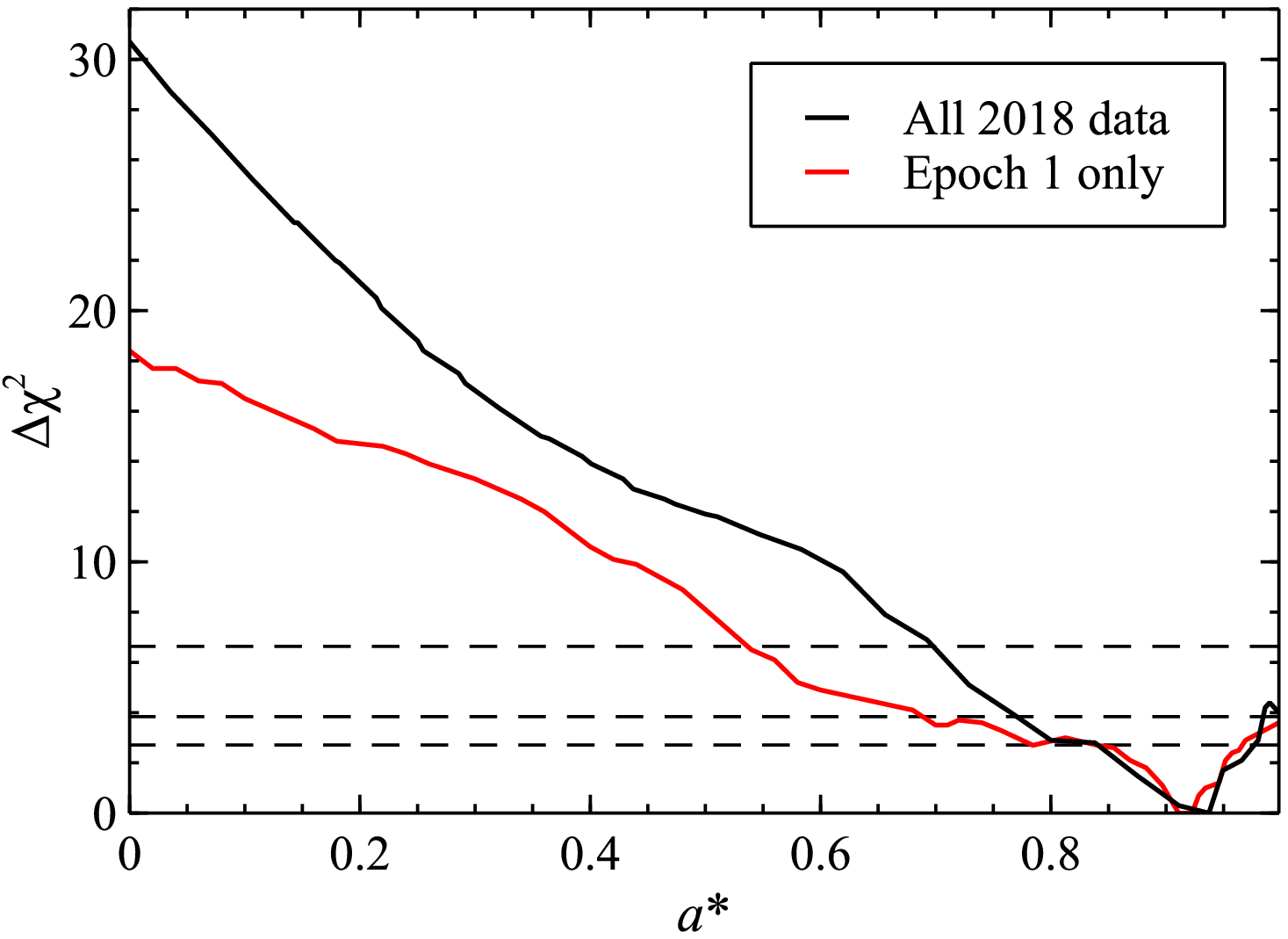}}
}
\end{center}
\vspace*{-0.3cm}
\caption{
The $\Delta$\chisq\ confidence contours for the spin of \iras\ based on our spectral
modeling of the full 2018 dataset with \Rfrac\ computed self-consistently from $a^*$ and
$h$ in the lamppost geometry (black; see Section \ref{sec_2018}). For comparison, we
also show the equivalent contour based on just the coordinated \xmm+\nustar\ data
(epoch 1; red, as in Figure \ref{fig_spin_XN1}). The horizontal dotted lines again
represent the 90, 95 and 99\% confidence levels for a single parameter of interest.}
\label{fig_spin_2018}
\end{figure}

The final fit to the full 2018 dataset is again very good, with \chisq/DoF = 4749/4583.
We give the constraints on the variable model parameters in Table \ref{tab_param_2018};
the best-fit is still extremely similar to that found for epoch 1 alone. As such, we just show
the data/model ratio for the additional datasets (epochs 2a and 2b) in Figure
\ref{fig_ratio_e2}. We also compute the observed and absorption-corrected fluxes for the
full model and the \relxill\ component, respectively (Table \ref{tab flux}), to further highlight
the variability accounted for by the model. For epoch 2a, the photon index is slightly
harder than epochs 1 and 2b, and the column densities of the warm absorber components
also show changes in the relative contributions of the two components: there is now a
larger column of lower ionisation material (WA1) and a smaller column of higher ionisation
material (WA2) along our line-of-sight to the central nucleus. Although there appear to be
changes in both the intrinsic continuum and the line-of-sight absorption properties, the
change in spectral hardness seen during epoch 2a is primarily driven by the intrinsic
continuum. We note that linking the WA column densities across all epochs, such that the
WA components are completely stable, only results in a mild degradation in the fit
($\Delta\chi^{2} = 12$ for 2 fewer free parameters), and does not change any of the key
inner disc reflection parameters of interest here (\eg\ $a^*, i$). The outflow velocity of the
HIA has also decreased between epochs 1 and 2 (although this naturally has little effect
on the overall hardness of the spectra). Forcing the outflow velocity to be the same for
both epochs results in a significantly worse fit ($\Delta\chi^{2} = 18$ for 1 less free
parameter). To provide the most robust constraints we re-compute the confidence contour
for the black hole spin with this joint fit, and compare these with the constraints from epoch
1 in Figure \ref{fig_spin_2018}. The formal 90\% constraints are still similar and low spin
values are excluded at a much higher level of confidence than with the epoch 1 data only.

\section{Discussion}
\label{sec_dis}

We have presented a detailed analysis of the 2018 broadband X-ray observations of
the type 1 Seyfert \iras, combining \xmm, \nustar\ and \swift. The observed X-ray
spectrum is complex; the low energies are heavily influenced by a partially-ionised
`warm' absorber (\ovii/\oviii\ absorption edges, as found previously by \citealt{Ricci17}),
while the higher energies show clear evidence for strong relativistic reflection from the
inner accretion disc (relativistically broadened iron emission and associated Compton
reflection continuum, as tentatively suggested by \citealt{Liebmann18}). There is also
evidence for more distant reprocessing (narrow iron emission) and absorption by more
highly-ionised material (\fexxv/\fexxvi\ absorption) The broadband coverage provided by
\xmm, \nustar\ and \swift\ allows us to robustly disentangle these various effects.

These data span two epochs, and flux variability is clearly observed, along with some
moderate spectral variability. Although we find evidence that the properties of the warm
absorber are variable to some degree, the majority of the observed variability appears
to be driven by changes intrinsic to the source, and in particular the properties of the
primary Comptonised X-ray continuum (\ie $\Gamma$ and intrinsic source flux).

\subsection{Black Hole Mass}

The X-ray variability is characterised in Section \ref{sec_var}, and we find evidence for
a fairly standard AGN PSD (see Figure \ref{fig_psd}): roughly flat-topped at lower
frequencies before breaking to a steep spectrum at higher frequencies (\eg\
\citealt{Uttley02, Markowitz03, Papadakis10, GonzalezMartin12, Alston19iras}). We note
that the temporal separation of the two observing epochs was particularly useful in
constraining the low-frequency part of the PSD here. The detection of the PSD break
frequency provides us with an opportunity to constrain the mass of the black hole in \iras\
(\citealt{McHardy06, GonzalezMartin12}), provided the bolometric luminosity, \lbol, is also
known. As discussed in Section \ref{sec_var}, based on our CARMA modelling of the
PSD, we adopt a break timescale of $T_{\rm{b}} = 1.0 \pm 0.5$ days.

We estimate \lbol\ from the intrinsic (\ie absorption corrected) 2--10\,keV luminosities
calculated for the \relxill\ component from our spectral fits to the broadband data, utilizing
the available 2--10\,keV bolometric corrections in the literature ($\kappa_{2-10} \equiv
L_{\rm{bol}}/L_{2-10}$). For a luminosity distance of $D = 267$\,Mpc (assuming a
standard $\Lambda$CDM concordance cosmology, \ie \lambdaCDM), the rest-frame
2--10\,keV luminosities for epochs 1--2 are $L_{2-10} = 1.6-1.9 \times 10^{44}$\,\ergps.
Given the source spent more time at the higher end of this flux range during our
observations, we adopt an average 2--10\,keV luminosity of $L_{2-10} = 1.8 \times
10^{44}$\,\ergps\ for the 2018 dataset. The appropriate value of $\kappa_{2-10}$
depends on the Eddington ratio, $\lambda_{\rm{E}} \equiv L_{\rm{bol}}/L_{\rm{Edd}}$;
$\kappa_{2-10}$ varies from $\sim$10 for $\lambda_{\rm{E}} \lesssim 0.01$ up to
$\sim$100 for $\lambda_{\rm{E}} \sim 1$ (\eg\ \citealt{Vasudevan09, Lusso10}). We
estimate $\lambda_{\rm{E}}$ from the known correlation between $\lambda_{\rm{E}}$
and the X-ray photon index (\eg\ \citealt{Shemmer08, Risaliti09LxGam, Brightman13}).
Based on the most recent of these works (\citealt{Brightman13}), the values of
$\Gamma$ found here ($2.07-2.16$) imply an Eddington fraction of $\lambda_{\rm{E}}
\sim 0.4$. In turn, this implies a bolometric correction of $\kappa_{2-10} \sim 50$, and a
bolometric luminosity of $L_{\rm{bol}} \sim 9 \times 10^{45}$\,\ergps. Given the scatter
seen in $\kappa_{2-10} \sim 50$ (\citealt{Lusso10}), we estimate the uncertainty on this
value to be at least a factor of $\sim$2.

Combining this with the break timescale from the PSD, we estimate a black hole
mass of $\log[M_{\rm{BH}}/M_{\odot}] = 8.0 \pm 0.6$ from the relation linking
$M_{\rm{BH}}$, \lbol\ and $T_{\rm{b}}$ presented by \cite{McHardy06}.\footnote{Although
\cite{GonzalezMartin12} formally present a more recent evaluation of the connection
between $M_{\rm{BH}}$, \lbol\ and $T_{\rm{b}}$, we use the original \cite{McHardy06}
work here for two reasons. First, these more recent works are based on the PSD
properties calculated across the full \xmm\ band (0.3--10.0\,keV), while the
\cite{McHardy06} work is based on the 2--10\,keV band, which is a much better match to
the \nustar\ bandpass (we use the 3--10\,keV band for our PSD analysis). Although any
energy dependence in the break frequency is expected to be subtle for AGN (\eg\
\citealt{Alston19iras}), we feel it best to err on the side of caution here. Second, we have
some concerns about the sample selection in the recent evaluation. Most notably, the
sample on which this is based includes the Circinus nucleus despite this being one of the
best-known Compton-thick AGN (\eg\ \citealt{Matt96circ, Bianchi02circ, Arevalo14}),
meaning the intrinsic AGN continuum is not seen below 10\,keV. The variability seen by
\xmm\ is instead almost certainly related to the bright X-ray binaries that are within the
$40''$ extraction region used by \cite{GonzalezMartin12}, which make a significant
contribution to the total soft X-ray emission; most notable is the variable ultraluminous
X-ray source CG\,X-1 which can reach luminosities in excess of $10^{40}$\,\ergps\ and is
separated from the nucleus by $\sim$15$''$ (\citealt{Bauer01, Qiu19}). However, we
stress that the \cite{GonzalezMartin12} evaluation ultimately still agrees with the result
presented here, giving $\log[M_{\rm{BH}}/M_{\odot}] = 7.7 \pm 0.6$.} The final uncertainty
quoted here comes from combining (in quadrature) the estimated 1$\sigma$ uncertainties
on $T_{\rm{b}}$ and $L_{\rm{bol}}$ ($\sim$0.1 and $\sim$0.2\,dex, respectively) with the
uncertainty on the absolute mass calibration used when deriving the scaling relation
(taken to be $\sim$0.4\,dex; \citealt{Peterson14rev}). Although there is good consistency
between the values for \kbol210\ and \lbol\ obtained here and the equivalent values
obtained by \cite{Vasudevan10}, who estimated \lbol\ based on the infrared luminosity,
the black hole mass obtained here is significantly smaller than the mass presented in that
work, $M_{\rm{BH}} \sim 3 \times 10^9$\,\msun, estimated from the relation between
\mbh\ and the $K$-band bulge luminosity (\citealt{Marconi03}). However, as a sanity
check, we note that the Eddington ratio implied by our estimated mass and bolometric
luminosity is close to (but still below) unity, in reasonable agreement with that estimated
from $\Gamma$ (which did not necessarily need to have been the case). Furthermore,
the mass estimated here is in good agreement with that obtained by \cite{Parisi09} based
on the \hbeta\ line width, $\log[M_{\rm{BH}}/M_{\odot}] \sim 7.9$, and also with that
obtained by the BASS collaboration based on the \Ha\ line width,
$\log[M_{\rm{BH}}/M_{\odot}] \sim 8.4$ (\citealt{Koss17}). We are therefore satisfied that
our mass of $\log[M_{\rm{BH}}/M_{\odot}] = 8.0 \pm 0.6$ is robust\footnote{\textit{Note
added after acceptance:} the mass obtained here is also in excellent agreement with that
very recently posted by the GRAVITY collaboration (\citealt{Grav20iras09149}), and we
note that there is also good consistency between the inclination they find for the broad line
region (for both of the models presented) and the inclination we obtain for the inner
accretion disc.}, and provides a self-consistent solution for \iras. The discrepancy with the
mass reported in \cite{Vasudevan09} is likely because \iras\ is still AGN-dominated in the
$K$-band, given that it shows clear broad emission lines in the optical (\citealt{Perez89}),
resulting in an overestimate of the luminosity of the bulge and in turn the black hole mass.

\subsection{Black Hole Spin}

In addition to the mass constraint from the X-ray variability, the relativistic reflection
features in the broadband X-ray spectrum allow us to measure the spin. We model this
reflection self-consistently in the context of the lamppost geometry, which we find
provides a very good description of the data despite being a clearly simplified geometry
(\eg\ \citealt{Wilkins12, Zhang19}), including correctly predicting the observed reflection
fraction (see Figure \ref{fig_Rfrac}). This is further support for the idea that the X-ray
corona is compact and centrally located. To provide the most robust constraint on the
spin, our final analysis is based on a joint fit to all of the 2018 data (epochs 1 and 2); we
find that \iras\ hosts a rapidly rotating black hole with $a^* = 0.94^{+0.02}_{-0.07}$ (see
Figure \ref{fig_spin_2018}). These observations have therefore allowed us to fully
characterise the supermassive black hole in \iras. 

Systematic uncertainties on spin measurements from reflection analyses are difficult to
quantify, but are likely $\Delta a^* \sim 0.1$ for rapidly rotating black holes with strong
reflection (\eg\ \citealt{Bonson16, Choudhury17, Kammoun18}), \ie similar to the
statistical uncertainty in this case. We note that the models used here assume that the
disc is essentially razor thin; this may be one source of systematic error, as in reality the
disc is likely to have some non-negligible vertical extent. For an Eddington ratio of
$\lambda_{\rm{E}} \sim 0.4$, as inferred above, the maximum scale-height of the disc
should be $H/R \sim 0.15$ (\citealt{McClintock06}). Significant `bleeding' of the reflected
emission over the ISCO is therefore unlikely (\citealt{Reynolds08}), and we are also
unlikely to have introduced significant uncertainties by assuming an emissivity profile for
a thin disc (and if anything, the latter would cause us to underestimate the spin;
\citealt{fenrir}).

The spin constraint from this analysis comes from modelling the full suite of reflection
features present in our broadband spectral model, including the relativistically
broadened iron emission, the strength of the Compton reflection continuum, and the
soft excess (which is partially seen through the ionised absorption; see Figure
\ref{fig_model}). Although the first two features can be readily seen in the broadband
data, the presence of the ionised absorption makes it challenging to unambiguously
test whether a soft excess is present in \iras. Nevertheless, is seen almost ubiquitously
in similar AGN that have low levels of obscuration, and its presence is implicitly
assumed in the broadband reflection modelling undertaken here. However, it is
important to note that the nature of the soft excess is still hotly debated. As implied
here, a reflection origin is often invoked (\eg\ \citealt{Crummy06, Walton13spin,
Jiang19agn}), in which the forest of low-energy fluorescent emission lines in the
rest-frame reflection spectrum are all relativistically broadened in the same way as the
iron emission, and blend together to form a smooth low-energy excess. This is
supported by the discovery that the soft excess exhibits the time lags relative to the
primary powerlaw continuum expected in this scenario (\eg\ \citealt{FabZog09,
Fabian13iras, deMarco13, Alston14, Alston20iras}); these lags are well explained by
reverberation of the inner disc, and are similar in amplitude to the lags seen from the
broad iron line (which is unambiguously associated with reflection from this region;
\citealt{Zoghbi12, Kara13feK, Kara15}). However, in some cases the reflection model
does not appear to fit the broadband data well (\eg\ \citealt{Matt14, Porquet18}), and an
alternative model invoking distinct Comptonizing zones for the soft excess (the `warm'
corona) and the primary powerlaw continuum (the `hot' corona) is frequently proposed
as an alternative (\eg\ \citealt{Done12, Petrucci13, Petrucci18, Middei19}; see
\citealt{Garcia19} and \citealt{Petrucci20} for the latest debate over whether such warm
coronae are physically plausible). Given this, we tested how sensitive the spin
constraints were to the treatment of the soft X-ray data (keeping the warm absorber
components fixed at their best-fit values from the full band analysis, since these cannot
be constrained with the data above 2\,keV). Based on the broadband data from epoch
1, we find the constraints are practically identical when fitting the data only above
2\,keV, \ie excluding the contribution of the soft excess (see Figure \ref{fig_spin_XN1}).
Our conclusion that \iras\ hosts a rapidly rotating black hole is therefore robust to the
precise nature of the soft excess.

Throughout this work we have made use of reflection models that assume the accretion
disc has a fixed electron density of $n_{\rm{e}} = 10^{15}$\,\pcmcub. This density has
been adopted as standard for the majority of the reflection models discussed in the
literature (\eg\ \citealt{Ross99, cdid, reflion, xillver}), and is motivated by the expected
value for a `typical' AGN, \ie a $\sim$10$^{8}$\,\msun\ black hole accreting at a
significant fraction of its Eddington luminosity (\eg\ \citealt{Svensson94}). However, while
it has been known for some time that this is not appropriate for the accretion discs around
X-ray binaries, which should have much higher densities (\eg\ \citealt{Reis09spin,
Walton12xrbAGN, King14, Tomsick18cyg, Jiang19gx}), more recently the density of the
disc has also been shown to be an important issue even for reflection modelling within
the AGN population, given the broad range of central black hole masses and accretion
rates observed (\citealt{Garcia16, Jiang18iras, Jiang19agn}). Larger densities increase
the rate of free-free absorption, resulting in significant changes in the reflection continuum
at low energies which can be important to account for, particularly when modelling the
soft excess. However, these effects can also influence the reflection continuum in the Fe
K band, and thus influence the iron abundance inferred (\citealt{Tomsick18cyg,
Jiang19agn}), so they are potentially important to consider in a broadband context as well.
Nevertheless, the `typical' AGN described above is very close to the scenario we infer for
\iras, so the density assumed in the models used here is actually a suitable choice, and
should not introduce any significant systematic uncertainties in our spin measurement.
Indeed, if we replace \relxilllpioncp\ with \relxilllpd\ in our analysis of the \xmm+\nustar\
data from epoch 1, allowing the disc density to be varied as a free parameter instead of
the radial ionisation gradient (it is not currently possible to vary both simultaneously with
the \relxill\ models), we find that $\log[n_{\rm{e}}/\rm{cm}^{-3}] < 15.2$. As expected, the
spin constraint is essentially unchanged.

We can therefore add \iras\ to the growing list of rapidly rotating black holes powering
radio-quiet AGN (\eg\ \citealt{Brenneman11, Gallo13, Risaliti13nat, Walton14,
Marinucci14mcg6, Svoboda15, Buisson18}). This is further evidence that, while black hole
spin may well play a significant role in powering the relativistic jets launched by accreting
black holes (\citealt{BZ77}), the angular momentum of the black hole can not be the only
ingredient necessary for jet launching (\citealt{King13jet}). The distinction between
radio-loud and radio-quiet AGN therefore cannot be simply driven by differences in spin,
as has previously been suggested (\eg\ \citealt{Wilson95, Moderski98, Sikora07}); this
would require that radio-quiet AGN host slowly rotating black holes, contrary to observation. 

The high spin obtained here also has implications for the most recent period of significant
growth experienced by the SMBH in \iras. This likely occurred via prolonged `coherent'
accretion (\ie the accreted material always has a common angular momentum axis), as
this tends to produce rapidly rotating black holes, while more chaotic accretion would
instead tend to spin the black hole down (\eg\ \citealt{Dubois14, Sesana14, Fiacconi18}).
There is growing evidence for a `top-heavy' spin distribution among local AGN (\ie high
spins are preferred; \citealt{Walton13spin, Reynolds14rev}), which would suggest that
such growth is common. However, caution is still required here, as there are known
selection biases towards observing high-spin objects (\citealt{Brenneman11,
Vasudevan16}) which are likely significant. Larger samples of spin measurements to
overcome this bias, combined with efforts to track the redshift evolution of black hole spin
(\eg\ \citealt{Reis14nat, Reynolds14, Walton15lqso}) are required to properly constrain
SMBH growth models in a statistical sense.

\subsection{Ionised Absorption}

The low-energy spectrum observed from \iras\ is heavily modified by the effects of
absorption by partially ionised material, particularly the \ovii/\oviii\ edges at
$\sim$0.7--0.8\,keV, as previously suggested by \cite{Ricci17}. The best-fit model found
here prefers two absorption components for the warm absorber: a slightly lower ionisation
component with $\log[\xi/(\rm{erg}~\rm{cm}~\rm{s}^{-1})] \sim 1.1$ and a slightly higher
ionisation component with $\log[\xi/(\rm{erg}~\rm{cm}~\rm{s}^{-1})] \sim 2.0$. This is more
complex than the single-component absorption model used by \cite{Ricci17}, but we stress
again that this is likely related to the much lower S/N data available to them at the time.
Such complexity in the warm absorber is not unusual where high S/N data is available
(\eg\ \citealt{Lee01, Krongold03, Steenbrugge05, Reeves13}). The parameters we find
($\xi$, $N_{\rm{H}}$, $v_{\rm{out}}$) are fairly typical when compared against the warm
absorbers seen in other systems (\eg\ \citealt{Laha14}); the outflow velocities
($\sim$4000--7000\,\kmps) could be considered slightly on the high side, but are not
unprecedented for such absorption (\eg\ \citealt{Longinotti19}).

In addition to the warm absorber seen at low energies, we also see evidence for
absorption from much more highly ionised material in the iron band with
$\log[\xi/(\rm{erg}~\rm{cm}~\rm{s}^{-1})] \sim 3.5$ (giving \fexxv/\fexxvi\ absorption). We
find this to be the fastest outflowing component in epoch 1 ($v_{\rm{out}} \sim
9000$\,\kmps), although the velocity has dropped in epoch 2 ($v_{\rm{out}} \lesssim
5000$\,\kmps). Qualitatively similar stratification of the various outflowing zones (higher
velocity at higher ionisation) to that found in epoch 1 and velocity variability in other
highly ionised outflows have both been seen previously (\eg\ \citealt{Kosec18_1h0707,
Matzeu17, Pinto18iras}). Although this component appears to reach reasonably large 
outflow velocities, the outflow seen here still appears to be relatively slow in comparison
to the most extreme seen in other AGN (`ultrafast' outflows, which can reach velocities
of $\sim$0.4$c$; \citealt{Reeves18, Walton19ufo}). 

Taking the observed results at face-value, and following previous work (\eg\
\citealt{Nardini15, Walton19ufo}), we attempt to estimate the kinetic power, $L_{\rm{kin}}$,
of the highly ionised component relative to the bolometric radiative output via equation
\ref{eqn_wind}:

\begin{equation}
\frac{L_{\rm{kin}}}{L_{\rm{bol}}}\approx2{\pi}m_{\rm{p}}\mu\frac{R_{\rm{w}} N_{\rm{H}}
v_{\rm{out}}^3}{L_{\rm{bol}}}{\Omega}C_{\rm{V}}
\label{eqn_wind}
\end{equation}
\vspace*{0.1cm}

\noindent{where} $\mu$ is the mean atomic weight $\sim$1.2 for solar abundances,
$m_{\rm{p}}$ is the proton mass, $R_{\rm{w}}$ is the radius of the wind, and $\Omega$
and $C_{\rm{V}}$ are the unknown solid angle and volume filling factor of the absorber,
respectively (both normalised to vary between 0--1; note that $\Omega$ is formally distinct
from $C_{\rm{f}}$, which is the line-of-sight covering factor). While $R_{\rm{w}}$ is not
known here, we can set a lower limit on this ratio by taking this to be the escape radius
implied by the outflow velocity, \ie $R_{\rm{w}} = R_{\rm{esc}} = G M_{\rm{BH}} / 
v_{\rm{out}}^2$. This would imply that $L_{\rm{kin}}/L_{\rm{bol}} \gtrsim 3 \times 10^{-3} ~
\Omega C_{\rm{V}}$. Performing the same calculations for the WA1 and WA2 components
results in even smaller values of $L_{\rm{kin}}/L_{\rm{bol}}$ (although WA2 is of the same
order). Even assuming that these are all independent outflow components would therefore
only increase the total $L_{\rm{kin}}/L_{\rm{bol}}$ by a factor of $\sim$2.

Simulations predict that the winds launched by accretion discs should be largely equatorial
(\eg\ \citealt{Proga00, Proga04, Nomura16}); there is clear evidence that this is the case
for X-ray binaries (\citealt{Ponti12}), and there is also some evidence that AGN outflow
properties are inclination dependent (\citealt{Parker18ufo}). As the inclination inferred from
the reflection spectrum is fairly modest here, the true outflow velocity could yet be slightly
larger owing to projection effects. However, for $i \sim 40-45$\deg\ the intrinsic velocity
could only be up to a factor of $\sim$1.5 larger, and $L_{\rm{kin}}$ would only increase by
the same factor for $R_{\rm{w}} = R_{\rm{esc}}$ (since $R_{\rm{esc}} \propto
v_{\rm{out}}^{-2}$). We also note that, given the way they have been normalised here, the
product $\Omega C_{\rm{V}}$ must be $\leq 1$. Unless $R_{\rm{w}} \gg R_{\rm{esc}}$
then it is not clear the outflow seen here can be sufficient to drive galaxy-scale feedback;
simulations suggest that $L_{\rm{kin}}$ must be at least a few per cent of \lbol\ to do so
(\eg\ \citealt{DiMatteo05nat, Hopkins10}). A more powerful outflow may yet be present in
\iras, particularly given that we infer it is accreting at close to its Eddington limit. However,
if this is the case it either does not intercept our line-of-sight, which is plausible for a
viewing angle of $i \sim 40-45$\deg, or is too highly ionised for a significant detection.

\section{Conclusions}
\label{sec_conc}

Combining X-ray timing and spectroscopy, we have been able to fully characterise the
supermassive black hole in the type I Seyfert galaxy \iras, a complex source that has
received relatively little observational attention to date. We find the mass of the black
hole to be $\log[M_{\rm{BH}}/M_{\odot}] = 8.0 \pm 0.6$ (primarily from X-ray timing
constraints on the PSD break frequency provided by \nustar) and the spin of the black
hole to be $a^* = 0.94^{+0.02}_{-0.07}$ (from broadband X-ray spectroscopic constraints
on the relativistic reflection from the inner disc, combining \xmm, \nustar\ and \swift). The
mass obtained here is in good agreement with that estimated previously from the \hbeta\
line width, and implies that the black hole in \iras\ is accreting at a reasonable fraction of
its Eddington luminosity. The spin constraint presented here is the first available in the
literature for \iras, and shows this to be another example of a radio-quiet AGN powered
by a rapidly rotating black hole.

\section*{ACKNOWLEDGEMENTS}

The authors would like to thank the reviewer for the feedback provided, which helped
improve the final version of the manuscript.
DJW acknowledges support from the Science and Technology Facilities Council (STFC)
via an Ernest Rutherford Fellowship, and PK acknowledges support from an STFC
studentship.
JAG acknowledges support from NASA grants 80NSSC19K1020 and 80NSSC19K0586,
and from the Alexander von Humboldt Foundation.
EN acknowledges financial contribution from the agreement ASI-INAF n.2017-14-H.0 and
partial support from the EU Horizon 2020 Marie Sk\l{}odowska-Curie grant agreement no.
664931.
CR acknowledges support from the Fondecyt Iniciacion grant 11190831.
CSR thanks STFC for support under the New Applicant grant ST/R000867/1, and the
European Research Council for support under the European Union’s Horizon 2020
research and innovation programme (grant 834203). 
This research has made use of data obtained with \nustar, a project led by Caltech,
funded by NASA and managed by NASA Jet Propulsion Laboratory (JPL), and has utilized
the \nustardas\ software package, jointly developed by the Space Science Data Centre
(SSDC; Italy) and Caltech (USA).
This research has also made use of data obtained with \xmm, an ESA science mission
with instruments and contributions directly funded by ESA Member States, as well as
public data from the \swift\ data archive.
This work has made use of the \corner\ package (\citealt{corner}) for data visualisation.

\textit{\textbf{Data Availability Statement:}} The data underlying this article are publicly
available from ESA's \textit{XMM-Newton} Science Archive
(https://www.cosmos.esa.int/web/xmm-newton/xsa)  and NASA's HEASARC archive
(https://heasarc.gsfc.nasa.gov/).

\appendix

\section{Further Monte Carlo Results}
\label{app_mcmc}

\begin{figure*}
\begin{center}
\hspace*{-0.1cm}
\rotatebox{0}{
{\includegraphics[width=470pt]{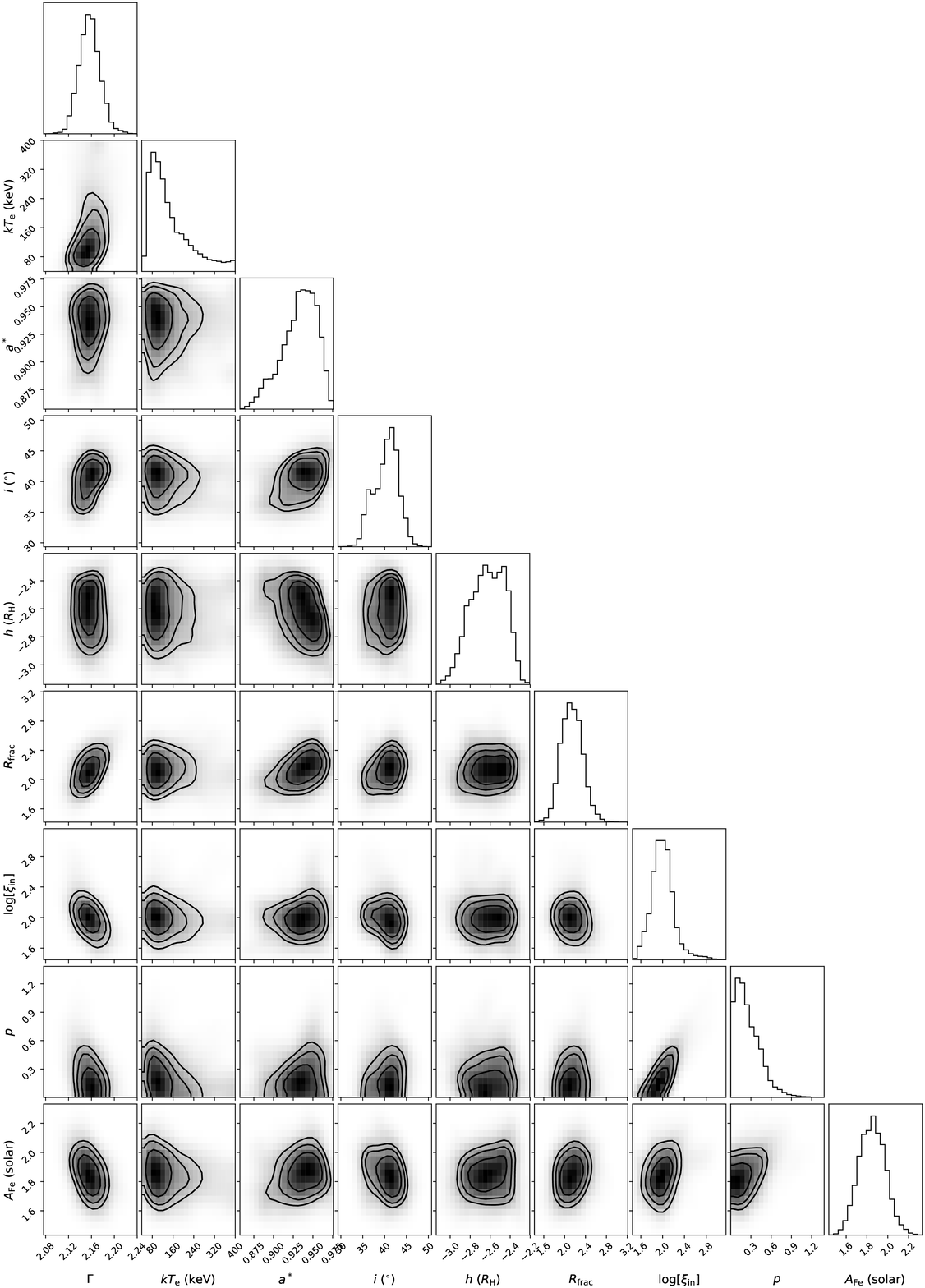}}
}
\hspace{0.6cm}
\end{center}
\vspace*{-0.3cm}
\caption{The MCMC results for parameters relating to the intrinsic AGN continuum and
the relativistic disc reflection for the data from epoch 1. The plot format follows that of
Figure \ref{fig_mcmc}.
}
\label{fig_app_mcmc1}
\end{figure*}

\begin{figure*}
\begin{center}
\hspace*{-0.1cm}
\rotatebox{0}{
{\includegraphics[width=470pt]{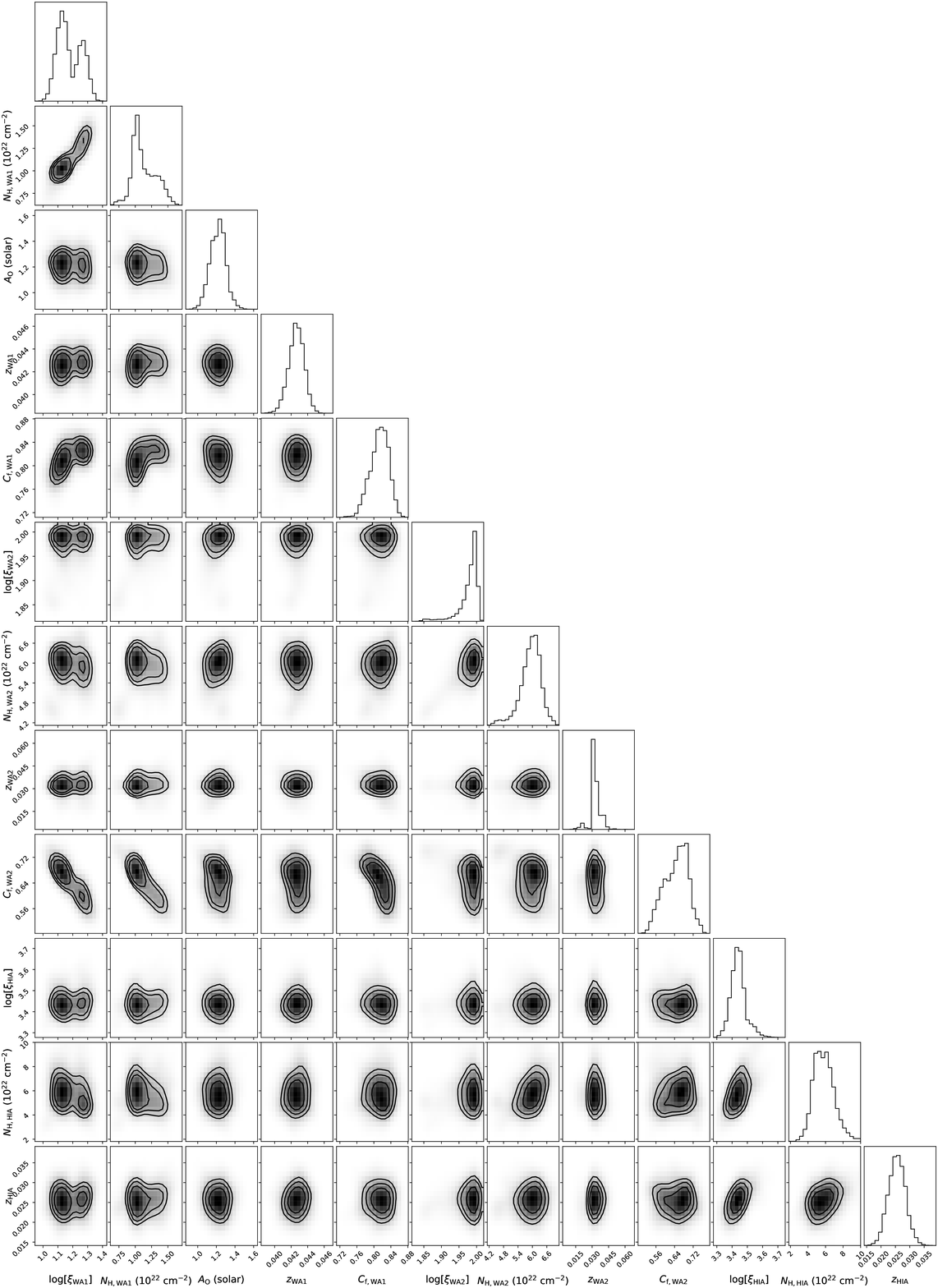}}
}
\hspace{0.6cm}
\end{center}
\vspace*{-0.3cm}
\caption{The MCMC results for parameters relating to the various ionised absorbers for
the data from epoch 1. The plot format again follows that of Figures \ref{fig_mcmc} and
\ref{fig_app_mcmc1}.
}
\label{fig_app_mcmc2}
\end{figure*}

Here we present the results from our MCMC simulations performed for the data from
epoch 1 (see Section \ref{sec_XN1}) for a variety of additional parameter combinations.
In Figure \ref{fig_app_mcmc1} we focus on the parameters relating to the intrinsic
continuum and the disc reflection, and in Figure \ref{fig_app_mcmc2} we focus on the
parameters relating to the various ionised absorbers. Note that here, the lamppost
height is in units of the vertical horizon (\rh, hence the negative values which relate to the
\relxill\ setup) which varies from $1 \leq R_{\rm{H}}/R_{\rm{G}} \leq 2$, depending on the
spin. In addition, outflow velocities for the absorbers are again given in terms of their
redshifts in the observed frame, as in Figure \ref{fig_mcmc}.

\bibliographystyle{/Users/dwalton/papers/mnras}

\bibliography{/Users/dwalton/papers/references}

\begin{thebibliography}{160}
\expandafter\ifx\csname natexlab\endcsname\relax\def\natexlab#1{#1}\fi

\bibitem[{Alston} et~al.(2014{\natexlab{a}}){Alston}, {Done} \&
  {Vaughan}]{Alston14}
{Alston} W.~N., {Done} C., {Vaughan} S., 2014{\natexlab{a}}, \mnras, 439, 2,
  1548

\bibitem[{Alston} et~al.(2019){Alston}, {Fabian}, {Buisson}
  et~al.]{Alston19iras}
{Alston} W.~N., {Fabian} A.~C., {Buisson} D.~J.~K., et~al., 2019, \mnras, 482,
  2, 2088

\bibitem[{Alston} et~al.(2020){Alston}, {Fabian}, {Kara} et~al.]{Alston20iras}
{Alston} W.~N., {Fabian} A.~C., {Kara} E., et~al., 2020, Nature Astronomy, ~2

\bibitem[{Alston} et~al.(2014{\natexlab{b}}){Alston}, {Markeviciute}, {Kara},
  {Fabian} \& {Middleton}]{Alston14qpo}
{Alston} W.~N., {Markeviciute} J., {Kara} E., {Fabian} A.~C., {Middleton} M.,
  2014{\natexlab{b}}, \mnras, 445, L16

\bibitem[{Ar{\'e}valo} et~al.(2014){Ar{\'e}valo}, {Bauer}, {Puccetti}
  et~al.]{Arevalo14}
{Ar{\'e}valo} P., {Bauer} F.~E., {Puccetti} S., et~al., 2014, \apj, 791, 81

\bibitem[{Arnaud}(1996)]{xspec}
{Arnaud} K.~A., 1996, in { Astronomical Data Analysis Software and Systems
  V\/}, edited by {G.~H.~Jacoby \& J.~Barnes}, vol. 101 of { Astron. Soc. Pac.
  Conference Series, Astron. Soc. Pac., San Francisco\/}, ~17

\bibitem[{Ballantyne} et~al.(2001){Ballantyne}, {Ross} \& {Fabian}]{cdid}
{Ballantyne} D.~R., {Ross} R.~R., {Fabian} A.~C., 2001, \mnras, 327, 10

\bibitem[{Bauer} et~al.(2001){Bauer}, {Brandt}, {Sambruna} et~al.]{Bauer01}
{Bauer} F.~E., {Brandt} W.~N., {Sambruna} R.~M., et~al., 2001, \aj, 122, 182

\bibitem[{Bentz} et~al.(2009){Bentz}, {Walsh}, {Barth} et~al.]{Bentz09}
{Bentz} M.~C., {Walsh} J.~L., {Barth} A.~J., et~al., 2009, \apj, 705, 199

\bibitem[{Bianchi} et~al.(2002){Bianchi}, {Matt}, {Fiore}, {Fabian}, {Iwasawa}
  \& {Nicastro}]{Bianchi02circ}
{Bianchi} S., {Matt} G., {Fiore} F., {Fabian} A.~C., {Iwasawa} K., {Nicastro}
  F., 2002, \aap, 396, 793

\bibitem[{Bird} et~al.(2007){Bird}, {Malizia}, {Bazzano} et~al.]{3ISGRI}
{Bird} A.~J., {Malizia} A., {Bazzano} A., et~al., 2007, \apjs, 170, 1, 175

\bibitem[{Blandford} \& {Znajek}(1977)]{BZ77}
{Blandford} R.~D., {Znajek} R.~L., 1977, \mnras, 179, 433

\bibitem[{Blustin} et~al.(2005){Blustin}, {Page}, {Fuerst},
  {Branduardi-Raymont} \& {Ashton}]{Blustin05}
{Blustin} A.~J., {Page} M.~J., {Fuerst} S.~V., {Branduardi-Raymont} G.,
  {Ashton} C.~E., 2005, \aap, 431, 111

\bibitem[{Bonson} \& {Gallo}(2016)]{Bonson16}
{Bonson} K., {Gallo} L.~C., 2016, \mnras, 458, 1927

\bibitem[{Brenneman} et~al.(2011){Brenneman}, {Reynolds}, {Nowak}
  et~al.]{Brenneman11}
{Brenneman} L.~W., {Reynolds} C.~S., {Nowak} M.~A., et~al., 2011, \apj, 736,
  103

\bibitem[{Brightman} et~al.(2013){Brightman}, {Silverman}, {Mainieri}
  et~al.]{Brightman13}
{Brightman} M., {Silverman} J.~D., {Mainieri} V., et~al., 2013, \mnras, 433,
  2485

\bibitem[{Buisson} et~al.(2018){Buisson}, {Parker}, {Kara} et~al.]{Buisson18}
{Buisson} D.~J.~K., {Parker} M.~L., {Kara} E., et~al., 2018, \mnras, 480, 3,
  3689

\bibitem[{Burrows} et~al.(2005){Burrows}, {Hill}, {Nousek} et~al.]{SWIFT_XRT}
{Burrows} D.~N., {Hill} J.~E., {Nousek} J.~A., et~al., 2005, Space Science
  Reviews, 120, 165

\bibitem[{Cappi} et~al.(2016){Cappi}, {De Marco}, {Ponti} et~al.]{Cappi16}
{Cappi} M., {De Marco} B., {Ponti} G., et~al., 2016, \aap, 592, A27

\bibitem[{Choudhury} et~al.(2017){Choudhury}, {Garc{\'\i}a}, {Steiner} \&
  {Bambi}]{Choudhury17}
{Choudhury} K., {Garc{\'\i}a} J.~A., {Steiner} J.~F., {Bambi} C., 2017, \apj,
  851, 1, 57

\bibitem[{Cram} et~al.(1992){Cram}, {North} \& {Savage}]{Cram92}
{Cram} L.~E., {North} A., {Savage} A., 1992, \mnras, 257, 602

\bibitem[{Crummy} et~al.(2006){Crummy}, {Fabian}, {Gallo} \& {Ross}]{Crummy06}
{Crummy} J., {Fabian} A.~C., {Gallo} L., {Ross} R.~R., 2006, \mnras, 365, 1067

\bibitem[{Dauser} et~al.(2016){Dauser}, {Garc{\'{\i}}a}, {Walton}
  et~al.]{relxill_norm}
{Dauser} T., {Garc{\'{\i}}a} J., {Walton} D.~J., et~al., 2016, \aap, 590, A76

\bibitem[{De Marco} et~al.(2013){De Marco}, {Ponti}, {Cappi} et~al.]{deMarco13}
{De Marco} B., {Ponti} G., {Cappi} M., et~al., 2013, \mnras, 431, 2441

\bibitem[{den Herder} et~al.(2001){den Herder}, {Brinkman}, {Kahn}
  et~al.]{XMM_RGS}
{den Herder} J.~W., {Brinkman} A.~C., {Kahn} S.~M., et~al., 2001, \aap, 365, L7

\bibitem[{Di Matteo} et~al.(2005){Di Matteo}, {Springel} \&
  {Hernquist}]{DiMatteo05nat}
{Di Matteo} T., {Springel} V., {Hernquist} L., 2005, \nat, 433, 604

\bibitem[{Done} et~al.(2012){Done}, {Davis}, {Jin}, {Blaes} \& {Ward}]{Done12}
{Done} C., {Davis} S.~W., {Jin} C., {Blaes} O., {Ward} M., 2012, \mnras, 420,
  1848

\bibitem[{Dubois} et~al.(2014){Dubois}, {Volonteri} \& {Silk}]{Dubois14}
{Dubois} Y., {Volonteri} M., {Silk} J., 2014, \mnras, 440, 1590

\bibitem[{Fabian}(2012)]{Fabian12rev}
{Fabian} A.~C., 2012, \araa, 50, 455

\bibitem[{Fabian} et~al.(2013){Fabian}, {Kara}, {Walton} et~al.]{Fabian13iras}
{Fabian} A.~C., {Kara} E., {Walton} D.~J., et~al., 2013, \mnras, 429, 2917

\bibitem[{Fabian} et~al.(2009){Fabian}, {Zoghbi}, {Ross} et~al.]{FabZog09}
{Fabian} A.~C., {Zoghbi} A., {Ross} R.~R., et~al., 2009, \nat, 459, 540

\bibitem[{Ferrarese} et~al.(2006){Ferrarese}, {C{\^o}t{\'e}}, {Dalla Bont{\`a}}
  et~al.]{Ferrarese06}
{Ferrarese} L., {C{\^o}t{\'e}} P., {Dalla Bont{\`a}} E., et~al., 2006, \apjl,
  644, L21

\bibitem[{Fiacconi} et~al.(2018){Fiacconi}, {Sijacki} \& {Pringle}]{Fiacconi18}
{Fiacconi} D., {Sijacki} D., {Pringle} J.~E., 2018, \mnras

\bibitem[{Foreman-Mackey}(2016)]{corner}
{Foreman-Mackey} D., 2016, The Journal of Open Source Software, 1, 24

\bibitem[{Gallo} \& {Fabian}(2011)]{Gallo11}
{Gallo} L.~C., {Fabian} A.~C., 2011, \mnras, 418, L59

\bibitem[{Gallo} et~al.(2013){Gallo}, {Fabian}, {Grupe} et~al.]{Gallo13}
{Gallo} L.~C., {Fabian} A.~C., {Grupe} D., et~al., 2013, \mnras, 428, 1191

\bibitem[{Garc{\'{\i}}a} et~al.(2014){Garc{\'{\i}}a}, {Dauser}, {Lohfink}
  et~al.]{relxill}
{Garc{\'{\i}}a} J., {Dauser} T., {Lohfink} A., et~al., 2014, \apj, 782, 76

\bibitem[{Garc{\'{\i}}a} \& {Kallman}(2010)]{xillver}
{Garc{\'{\i}}a} J., {Kallman} T.~R., 2010, \apj, 718, 695

\bibitem[{Garc{\'\i}a} et~al.(2016){Garc{\'\i}a}, {Fabian}, {Kallman}
  et~al.]{Garcia16}
{Garc{\'\i}a} J.~A., {Fabian} A.~C., {Kallman} T.~R., et~al., 2016, \mnras,
  462, 1, 751

\bibitem[{Garc{\'\i}a} et~al.(2019){Garc{\'\i}a}, {Kara}, {Walton}
  et~al.]{Garcia19}
{Garc{\'\i}a} J.~A., {Kara} E., {Walton} D., et~al., 2019, \apj, 871, 1, 88

\bibitem[{Gehrels} et~al.(2004){Gehrels}, {Chincarini}, {Giommi} et~al.]{SWIFT}
{Gehrels} N., {Chincarini} G., {Giommi} P., et~al., 2004, \apj, 611, 1005

\bibitem[Geweke(1992)]{Geweke92}
Geweke J., 1992, in { Bayesian Statistics 4 (ed. J. M. Bernardo, J. O. Berger,
  A. P. Dawid and A. F. M. Smith)\/},  169--193, Clarendon Press, Oxford

\bibitem[{Gonz{\'a}lez-Mart{\'\i}n} \& {Vaughan}(2012)]{GonzalezMartin12}
{Gonz{\'a}lez-Mart{\'\i}n} O., {Vaughan} S., 2012, \aap, 544, A80

\bibitem[{Goodman} \& {Weare}(2010)]{MCMC_GW}
{Goodman} J., {Weare} J., 2010, Communications in Applied Mathematics and
  Computational Science, 5, 1, 65

\bibitem[{GRAVITY Collaboration} et~al.(2020){GRAVITY Collaboration}, {Amorim},
  {Brandner} et~al.]{Grav20iras09149}
{GRAVITY Collaboration}, {Amorim} A., {Brandner} W., et~al., 2020, arXiv
  e-prints,  arXiv:2009.08463

\bibitem[{Grevesse} \& {Sauval}(1998)]{Grevesse98}
{Grevesse} N., {Sauval} A.~J., 1998, \ssr, 85, 161

\bibitem[{Haardt} \& {Maraschi}(1991)]{Haardt91}
{Haardt} F., {Maraschi} L., 1991, \apjl, 380, L51

\bibitem[{Harrison} et~al.(2013){Harrison}, {Craig}, {Christensen}
  et~al.]{NUSTAR}
{Harrison} F.~A., {Craig} W.~W., {Christensen} F.~E., et~al., 2013, \apj, 770,
  103

\bibitem[{HI4PI Collaboration} et~al.(2016){HI4PI Collaboration}, {Ben Bekhti},
  {Fl{\"o}er} et~al.]{NH2016}
{HI4PI Collaboration}, {Ben Bekhti} N., {Fl{\"o}er} L., et~al., 2016, \aap,
  594, A116

\bibitem[{Hlavacek-Larrondo} et~al.(2012){Hlavacek-Larrondo}, {Fabian}, {Edge}
  et~al.]{HlavacekLarrondo12}
{Hlavacek-Larrondo} J., {Fabian} A.~C., {Edge} A.~C., et~al., 2012, \mnras,
  421, 2, 1360

\bibitem[{Hopkins} \& {Elvis}(2010)]{Hopkins10}
{Hopkins} P.~F., {Elvis} M., 2010, \mnras, 401, 7

\bibitem[{Ingram} et~al.(2019){Ingram}, {Mastroserio}, {Dauser}, {Hovenkamp},
  {van der Klis} \& {Garc{\'\i}a}]{Ingram19}
{Ingram} A., {Mastroserio} G., {Dauser} T., {Hovenkamp} P., {van der Klis} M.,
  {Garc{\'\i}a} J.~A., 2019, \mnras, 488, 1, 324

\bibitem[{Ishibashi} et~al.(2014){Ishibashi}, {Auger}, {Zhang} \&
  {Fabian}]{Ishibashi14}
{Ishibashi} W., {Auger} M.~W., {Zhang} D., {Fabian} A.~C., 2014, \mnras, 443,
  2, 1339

\bibitem[{Ishibashi} \& {Fabian}(2015)]{Ishibashi15}
{Ishibashi} W., {Fabian} A.~C., 2015, \mnras, 451, 1, 93

\bibitem[{Jansen} et~al.(2001){Jansen}, {Lumb}, {Altieri} et~al.]{XMM}
{Jansen} F., {Lumb} D., {Altieri} B., et~al., 2001, \aap, 365, L1

\bibitem[{Jiang} et~al.(2019{\natexlab{a}}){Jiang}, {Fabian}, {Dauser}
  et~al.]{Jiang19agn}
{Jiang} J., {Fabian} A.~C., {Dauser} T., et~al., 2019{\natexlab{a}}, \mnras,
  489, 3, 3436

\bibitem[{Jiang} et~al.(2019{\natexlab{b}}){Jiang}, {Fabian}, {Wang}
  et~al.]{Jiang19gx}
{Jiang} J., {Fabian} A.~C., {Wang} J., et~al., 2019{\natexlab{b}}, \mnras, 484,
  2, 1972

\bibitem[{Jiang} et~al.(2018){Jiang}, {Parker}, {Fabian} et~al.]{Jiang18iras}
{Jiang} J., {Parker} M.~L., {Fabian} A.~C., et~al., 2018, \mnras

\bibitem[{Kallman} \& {Bautista}(2001)]{xstar}
{Kallman} T., {Bautista} M., 2001, \apjs, 133, 221

\bibitem[{Kammoun} et~al.(2019){Kammoun}, {Dom{\v{c}}ek}, {Svoboda},
  {Dov{\v{c}}iak} \& {Matt}]{Kammoun19}
{Kammoun} E.~S., {Dom{\v{c}}ek} V., {Svoboda} J., {Dov{\v{c}}iak} M., {Matt}
  G., 2019, \mnras, 485, 1, 239

\bibitem[{Kammoun} et~al.(2018){Kammoun}, {Nardini} \& {Risaliti}]{Kammoun18}
{Kammoun} E.~S., {Nardini} E., {Risaliti} G., 2018, \aap, 614, A44

\bibitem[{Kara} et~al.(2013){Kara}, {Fabian}, {Cackett}, {Uttley}, {Wilkins} \&
  {Zoghbi}]{Kara13feK}
{Kara} E., {Fabian} A.~C., {Cackett} E.~M., {Uttley} P., {Wilkins} D.~R.,
  {Zoghbi} A., 2013, \mnras, 434, 1129

\bibitem[{Kara} et~al.(2015){Kara}, {Zoghbi}, {Marinucci} et~al.]{Kara15}
{Kara} E., {Zoghbi} A., {Marinucci} A., et~al., 2015, \mnras, 446, 737

\bibitem[{Kaspi} et~al.(2000){Kaspi}, {Smith}, {Netzer}, {Maoz}, {Jannuzi} \&
  {Giveon}]{Kaspi00}
{Kaspi} S., {Smith} P.~S., {Netzer} H., {Maoz} D., {Jannuzi} B.~T., {Giveon}
  U., 2000, \apj, 533, 631

\bibitem[{Kelly} et~al.(2014){Kelly}, {Becker}, {Sobolewska}, {Siemiginowska}
  \& {Uttley}]{CARMA}
{Kelly} B.~C., {Becker} A.~C., {Sobolewska} M., {Siemiginowska} A., {Uttley}
  P., 2014, \apj, 788, 1, 33

\bibitem[{King} et~al.(2013){King}, {Miller}, {G{\"u}ltekin} et~al.]{King13jet}
{King} A.~L., {Miller} J.~M., {G{\"u}ltekin} K., et~al., 2013, \apj, 771, 84

\bibitem[{King} et~al.(2014){King}, {Walton}, {Miller} et~al.]{King14}
{King} A.~L., {Walton} D.~J., {Miller} J.~M., et~al., 2014, \apjl, 784, L2

\bibitem[{Kormendy} \& {Ho}(2013)]{Kormendy13rev}
{Kormendy} J., {Ho} L.~C., 2013, \araa, 51, 511

\bibitem[{Kosec} et~al.(2018){Kosec}, {Buisson}, {Parker}, {Pinto}, {Fabian} \&
  {Walton}]{Kosec18_1h0707}
{Kosec} P., {Buisson} D.~J.~K., {Parker} M.~L., {Pinto} C., {Fabian} A.~C.,
  {Walton} D.~J., 2018, \mnras, 481, 1, 947

\bibitem[{Koss} et~al.(2017){Koss}, {Trakhtenbrot}, {Ricci} et~al.]{Koss17}
{Koss} M., {Trakhtenbrot} B., {Ricci} C., et~al., 2017, \apj, 850, 1, 74

\bibitem[{Krongold} et~al.(2003){Krongold}, {Nicastro}, {Brickhouse}, {Elvis},
  {Liedahl} \& {Mathur}]{Krongold03}
{Krongold} Y., {Nicastro} F., {Brickhouse} N.~S., {Elvis} M., {Liedahl} D.~A.,
  {Mathur} S., 2003, \apj, 597, 2, 832

\bibitem[{Laha} et~al.(2014){Laha}, {Guainazzi}, {Dewangan}, {Chakravorty} \&
  {Kembhavi}]{Laha14}
{Laha} S., {Guainazzi} M., {Dewangan} G.~C., {Chakravorty} S., {Kembhavi}
  A.~K., 2014, \mnras, 441, 3, 2613

\bibitem[{Lee} et~al.(2001){Lee}, {Ogle}, {Canizares} et~al.]{Lee01}
{Lee} J.~C., {Ogle} P.~M., {Canizares} C.~R., et~al., 2001, \apjl, 554, L13

\bibitem[{Liebmann} et~al.(2018){Liebmann}, {Fabian}, {Tsuruta}, {Haba} \&
  {Kunieda}]{Liebmann18}
{Liebmann} A.~C., {Fabian} A.~C., {Tsuruta} S., {Haba} Y., {Kunieda} H., 2018,
  \apj, 868, 1, 11

\bibitem[{Longinotti} et~al.(2019){Longinotti}, {Kriss}, {Krongold}
  et~al.]{Longinotti19}
{Longinotti} A.~L., {Kriss} G., {Krongold} Y., et~al., 2019, \apj, 875, 2, 150

\bibitem[{Lusso} et~al.(2010){Lusso}, {Comastri}, {Vignali} et~al.]{Lusso10}
{Lusso} E., {Comastri} A., {Vignali} C., et~al., 2010, \aap, 512, A34

\bibitem[{Lynden-Bell}(1969)]{LyndenBell69}
{Lynden-Bell} D., 1969, \nat, 223, 5207, 690

\bibitem[{Madsen} et~al.(2015){Madsen}, {Harrison}, {Markwardt}
  et~al.]{NUSTARcal}
{Madsen} K.~K., {Harrison} F.~A., {Markwardt} C.~B., et~al., 2015, \apjs, 220,
  8

\bibitem[{Malizia} et~al.(2007){Malizia}, {Landi}, {Bassani} et~al.]{Malizia07}
{Malizia} A., {Landi} R., {Bassani} L., et~al., 2007, \apj, 668, 1, 81

\bibitem[{Marconi} \& {Hunt}(2003)]{Marconi03}
{Marconi} A., {Hunt} L.~K., 2003, \apjl, 589, 1, L21

\bibitem[{Marinucci} et~al.(2014){Marinucci}, {Matt}, {Miniutti}
  et~al.]{Marinucci14mcg6}
{Marinucci} A., {Matt} G., {Miniutti} G., et~al., 2014, ArXiv 1404.3561

\bibitem[{Markowitz} et~al.(2003){Markowitz}, {Edelson}, {Vaughan}
  et~al.]{Markowitz03}
{Markowitz} A., {Edelson} R., {Vaughan} S., et~al., 2003, \apj, 593, 1, 96

\bibitem[{Matt} et~al.(1996){Matt}, {Fiore}, {Perola} et~al.]{Matt96circ}
{Matt} G., {Fiore} F., {Perola} G.~C., et~al., 1996, \mnras, 281, 4, L69

\bibitem[{Matt} et~al.(2014){Matt}, {Marinucci}, {Guainazzi} et~al.]{Matt14}
{Matt} G., {Marinucci} A., {Guainazzi} M., et~al., 2014, \mnras, 439, 3, 3016

\bibitem[{Matzeu} et~al.(2017){Matzeu}, {Reeves}, {Braito} et~al.]{Matzeu17}
{Matzeu} G.~A., {Reeves} J.~N., {Braito} V., et~al., 2017, \mnras, 472, L15

\bibitem[{McClintock} et~al.(2006){McClintock}, {Shafee}, {Narayan},
  {Remillard}, {Davis} \& {Li}]{McClintock06}
{McClintock} J.~E., {Shafee} R., {Narayan} R., {Remillard} R.~A., {Davis}
  S.~W., {Li} L.-X., 2006, \apj, 652, 518

\bibitem[{McHardy} et~al.(2006){McHardy}, {Koerding}, {Knigge}, {Uttley} \&
  {Fender}]{McHardy06}
{McHardy} I.~M., {Koerding} E., {Knigge} C., {Uttley} P., {Fender} R.~P., 2006,
  \nat, 444, 730

\bibitem[{Middei} et~al.(2018){Middei}, {Bianchi}, {Cappi} et~al.]{Middei18}
{Middei} R., {Bianchi} S., {Cappi} M., et~al., 2018, \aap, 615, A163

\bibitem[{Middei} et~al.(2019){Middei}, {Bianchi}, {Petrucci} et~al.]{Middei19}
{Middei} R., {Bianchi} S., {Petrucci} P.~O., et~al., 2019, \mnras, 483, 4, 4695

\bibitem[{Moderski} et~al.(1998){Moderski}, {Sikora} \& {Lasota}]{Moderski98}
{Moderski} R., {Sikora} M., {Lasota} J.-P., 1998, \mnras, 301, 142

\bibitem[{Moreno} et~al.(2019){Moreno}, {Vogeley}, {Richards} \&
  {Yu}]{Moreno19}
{Moreno} J., {Vogeley} M.~S., {Richards} G.~T., {Yu} W., 2019, \pasp, 131,
  1000, 063001

\bibitem[{Nardini} et~al.(2015){Nardini}, {Reeves}, {Gofford}
  et~al.]{Nardini15}
{Nardini} E., {Reeves} J.~N., {Gofford} J., et~al., 2015, Science, 347, 860

\bibitem[{Nomura} et~al.(2016){Nomura}, {Ohsuga}, {Takahashi}, {Wada} \&
  {Yoshida}]{Nomura16}
{Nomura} M., {Ohsuga} K., {Takahashi} H.~R., {Wada} K., {Yoshida} T., 2016,
  \pasj, 68, 1, 16

\bibitem[{Papadakis} et~al.(2010){Papadakis}, {Brinkmann}, {Gliozzi} \&
  {Raeth}]{Papadakis10}
{Papadakis} I.~E., {Brinkmann} W., {Gliozzi} M., {Raeth} C., 2010, \aap, 518,
  A28

\bibitem[{Parisi} et~al.(2009){Parisi}, {Masetti}, {Jim{\'e}nez-Bail{\'o}n}
  et~al.]{Parisi09}
{Parisi} P., {Masetti} N., {Jim{\'e}nez-Bail{\'o}n} E., et~al., 2009, \aap,
  507, 3, 1345

\bibitem[{Parker} et~al.(2018){Parker}, {Buisson}, {Jiang} et~al.]{Parker18ufo}
{Parker} M.~L., {Buisson} D.~J.~K., {Jiang} J., et~al., 2018, \mnras

\bibitem[{Parker} et~al.(2017){Parker}, {Pinto}, {Fabian} et~al.]{Parker17nat}
{Parker} M.~L., {Pinto} C., {Fabian} A.~C., et~al., 2017, \nat, 543, 83

\bibitem[{Perez} et~al.(1989){Perez}, {Manchado}, {Pottasch} \&
  {Garcia-Lario}]{Perez89}
{Perez} E., {Manchado} A., {Pottasch} S.~R., {Garcia-Lario} P., 1989, \aap,
  215, 262

\bibitem[{Peterson}(2014)]{Peterson14rev}
{Peterson} B.~M., 2014, \ssr, 183, 1-4, 253

\bibitem[{Peterson} et~al.(2004){Peterson}, {Ferrarese}, {Gilbert}
  et~al.]{Peterson04}
{Peterson} B.~M., {Ferrarese} L., {Gilbert} K.~M., et~al., 2004, \apj, 613, 2,
  682

\bibitem[{Petrucci} et~al.(2020){Petrucci}, {Gronkiewicz}, {Rozanska}
  et~al.]{Petrucci20}
{Petrucci} P.~O., {Gronkiewicz} D., {Rozanska} A., et~al., 2020, \aap, 634, A85

\bibitem[{Petrucci} et~al.(2013){Petrucci}, {Paltani}, {Malzac}
  et~al.]{Petrucci13}
{Petrucci} P.~O., {Paltani} S., {Malzac} J., et~al., 2013, \aap, 549, A73

\bibitem[{Petrucci} et~al.(2018){Petrucci}, {Ursini}, {De Rosa}
  et~al.]{Petrucci18}
{Petrucci} P.~O., {Ursini} F., {De Rosa} A., et~al., 2018, \aap, 611, A59

\bibitem[{Pinto} et~al.(2018){Pinto}, {Alston}, {Parker} et~al.]{Pinto18iras}
{Pinto} C., {Alston} W., {Parker} M.~L., et~al., 2018, \mnras, 476, 1021

\bibitem[{Ponti} et~al.(2012){Ponti}, {Fender}, {Begelman}, {Dunn}, {Neilsen}
  \& {Coriat}]{Ponti12}
{Ponti} G., {Fender} R.~P., {Begelman} M.~C., {Dunn} R.~J.~H., {Neilsen} J.,
  {Coriat} M., 2012, \mnras, 422, L11

\bibitem[{Porquet} et~al.(2018){Porquet}, {Reeves}, {Matt} et~al.]{Porquet18}
{Porquet} D., {Reeves} J.~N., {Matt} G., et~al., 2018, \aap, 609, A42

\bibitem[{Pounds} et~al.(2003){Pounds}, {Reeves}, {King}, {Page}, {O'Brien} \&
  {Turner}]{Pounds03}
{Pounds} K.~A., {Reeves} J.~N., {King} A.~R., {Page} K.~L., {O'Brien} P.~T.,
  {Turner} M.~J.~L., 2003, \mnras, 345, 705

\bibitem[{Proga} \& {Kallman}(2004)]{Proga04}
{Proga} D., {Kallman} T.~R., 2004, \apj, 616, 688

\bibitem[{Proga} et~al.(2000){Proga}, {Stone} \& {Kallman}]{Proga00}
{Proga} D., {Stone} J.~M., {Kallman} T.~R., 2000, \apj, 543, 686

\bibitem[{Qiu} et~al.(2019){Qiu}, {Soria}, {Wang} et~al.]{Qiu19}
{Qiu} Y., {Soria} R., {Wang} S., et~al., 2019, \apj, 877, 1, 57

\bibitem[{Reeves} et~al.(2020){Reeves}, {Braito}, {Chartas}, {Hamann}, {Laha}
  \& {Nardini}]{Reeves20}
{Reeves} J.~N., {Braito} V., {Chartas} G., {Hamann} F., {Laha} S., {Nardini}
  E., 2020, \apj, 895, 1, 37

\bibitem[{Reeves} et~al.(2018){Reeves}, {Braito}, {Nardini}, {Lobban}, {Matzeu}
  \& {Costa}]{Reeves18}
{Reeves} J.~N., {Braito} V., {Nardini} E., {Lobban} A.~P., {Matzeu} G.~A.,
  {Costa} M.~T., 2018, \apjl, 854, L8

\bibitem[{Reeves} et~al.(2013){Reeves}, {Porquet}, {Braito} et~al.]{Reeves13}
{Reeves} J.~N., {Porquet} D., {Braito} V., et~al., 2013, \apj, 776, 2, 99

\bibitem[{Reis} et~al.(2009){Reis}, {Fabian}, {Ross} \& {Miller}]{Reis09spin}
{Reis} R.~C., {Fabian} A.~C., {Ross} R.~R., {Miller} J.~M., 2009, \mnras, 395,
  1257

\bibitem[{Reis} et~al.(2014){Reis}, {Reynolds}, {Miller} \&
  {Walton}]{Reis14nat}
{Reis} R.~C., {Reynolds} M.~T., {Miller} J.~M., {Walton} D.~J., 2014, \nat,
  507, 207

\bibitem[{Reynolds}(1997)]{Reynolds97wa}
{Reynolds} C.~S., 1997, \mnras, 286, 3, 513

\bibitem[{Reynolds}(2014)]{Reynolds14rev}
{Reynolds} C.~S., 2014, \ssr, 183, 277

\bibitem[{Reynolds} \& {Fabian}(2008)]{Reynolds08}
{Reynolds} C.~S., {Fabian} A.~C., 2008, \apj, 675, 1048

\bibitem[{Reynolds} et~al.(2014){Reynolds}, {Walton}, {Miller} \&
  {Reis}]{Reynolds14}
{Reynolds} M.~T., {Walton} D.~J., {Miller} J.~M., {Reis} R.~C., 2014, \apjl,
  792, L19

\bibitem[{Ricci} et~al.(2017{\natexlab{a}}){Ricci}, {Trakhtenbrot}, {Koss}
  et~al.]{Ricci17}
{Ricci} C., {Trakhtenbrot} B., {Koss} M.~J., et~al., 2017{\natexlab{a}}, \apjs,
  233, 2, 17

\bibitem[{Ricci} et~al.(2017{\natexlab{b}}){Ricci}, {Trakhtenbrot}, {Koss}
  et~al.]{Ricci17nat}
{Ricci} C., {Trakhtenbrot} B., {Koss} M.~J., et~al., 2017{\natexlab{b}}, \nat,
  549, 7673, 488

\bibitem[{Risaliti} et~al.(2005){Risaliti}, {Bianchi}, {Matt}
  et~al.]{Risaliti05b}
{Risaliti} G., {Bianchi} S., {Matt} G., et~al., 2005, \apjl, 630, L129

\bibitem[{Risaliti} et~al.(2013){Risaliti}, {Harrison}, {Madsen}
  et~al.]{Risaliti13nat}
{Risaliti} G., {Harrison} F.~A., {Madsen} K.~K., et~al., 2013, \nat, 494, 449

\bibitem[{Risaliti} et~al.(2009){Risaliti}, {Young} \&
  {Elvis}]{Risaliti09LxGam}
{Risaliti} G., {Young} M., {Elvis} M., 2009, \apjl, 700, L6

\bibitem[{Ross} \& {Fabian}(2005)]{reflion}
{Ross} R.~R., {Fabian} A.~C., 2005, \mnras, 358, 211

\bibitem[{Ross} et~al.(1999){Ross}, {Fabian} \& {Young}]{Ross99}
{Ross} R.~R., {Fabian} A.~C., {Young} A.~J., 1999, \mnras, 306, 461

\bibitem[{Sesana} et~al.(2014){Sesana}, {Barausse}, {Dotti} \&
  {Rossi}]{Sesana14}
{Sesana} A., {Barausse} E., {Dotti} M., {Rossi} E.~M., 2014, \apj, 794, 104

\bibitem[{Shemmer} et~al.(2008){Shemmer}, {Brandt}, {Netzer}, {Maiolino} \&
  {Kaspi}]{Shemmer08}
{Shemmer} O., {Brandt} W.~N., {Netzer} H., {Maiolino} R., {Kaspi} S., 2008,
  \apj, 682, 1, 81

\bibitem[{Sikora} et~al.(2007){Sikora}, {Stawarz} \& {Lasota}]{Sikora07}
{Sikora} M., {Stawarz} {\L}., {Lasota} J.-P., 2007, \apj, 658, 2, 815

\bibitem[{Steenbrugge} et~al.(2005){Steenbrugge}, {Kaastra}, {Crenshaw}
  et~al.]{Steenbrugge05}
{Steenbrugge} K.~C., {Kaastra} J.~S., {Crenshaw} D.~M., et~al., 2005, \aap,
  434, 2, 569

\bibitem[{Str{\"u}der} et~al.(2001){Str{\"u}der}, {Briel}, {Dennerl}
  et~al.]{XMM_PN}
{Str{\"u}der} L., {Briel} U., {Dennerl} K., et~al., 2001, \aap, 365, L18

\bibitem[{Svensson} \& {Zdziarski}(1994)]{Svensson94}
{Svensson} R., {Zdziarski} A.~A., 1994, \apj, 436, 599

\bibitem[{Svoboda} et~al.(2015){Svoboda}, {Beuchert}, {Guainazzi},
  {Longinotti}, {Piconcelli} \& {Wilms}]{Svoboda15}
{Svoboda} J., {Beuchert} T., {Guainazzi} M., {Longinotti} A.~L., {Piconcelli}
  E., {Wilms} J., 2015, \aap, 578, A96

\bibitem[{Svoboda} et~al.(2012){Svoboda}, {Dov{\v{c}}iak}, {Goosmann}
  et~al.]{Svoboda12}
{Svoboda} J., {Dov{\v{c}}iak} M., {Goosmann} R.~W., et~al., 2012, \aap, 545,
  A106

\bibitem[{Taylor} \& {Reynolds}(2018)]{fenrir}
{Taylor} C., {Reynolds} C.~S., 2018, \apj, 855, 120

\bibitem[{Tombesi} et~al.(2010){Tombesi}, {Cappi}, {Reeves} et~al.]{Tombesi10b}
{Tombesi} F., {Cappi} M., {Reeves} J.~N., et~al., 2010, \aap, 521, A57

\bibitem[{Tomsick} et~al.(2018){Tomsick}, {Parker}, {Garc{\'{\i}}a}
  et~al.]{Tomsick18cyg}
{Tomsick} J.~A., {Parker} M.~L., {Garc{\'{\i}}a} J.~A., et~al., 2018, \apj,
  855, 3

\bibitem[{Tueller} et~al.(2008){Tueller}, {Mushotzky}, {Barthelmy}
  et~al.]{BAT9m}
{Tueller} J., {Mushotzky} R.~F., {Barthelmy} S., et~al., 2008, \apj, 681, 1,
  113

\bibitem[{Turner} et~al.(2001){Turner}, {Abbey}, {Arnaud} et~al.]{XMM_MOS}
{Turner} M.~J.~L., {Abbey} A., {Arnaud} M., et~al., 2001, \aap, 365, L27

\bibitem[{Uttley} et~al.(2002){Uttley}, {McHardy} \& {Papadakis}]{Uttley02}
{Uttley} P., {McHardy} I.~M., {Papadakis} I.~E., 2002, \mnras, 332, 1, 231

\bibitem[{Vasudevan} \& {Fabian}(2009)]{Vasudevan09}
{Vasudevan} R.~V., {Fabian} A.~C., 2009, \mnras, 392, 1124

\bibitem[{Vasudevan} et~al.(2010){Vasudevan}, {Fabian}, {Gandhi}, {Winter} \&
  {Mushotzky}]{Vasudevan10}
{Vasudevan} R.~V., {Fabian} A.~C., {Gandhi} P., {Winter} L.~M., {Mushotzky}
  R.~F., 2010, \mnras, 402, 1081

\bibitem[{Vasudevan} et~al.(2016){Vasudevan}, {Fabian}, {Reynolds}, {Aird},
  {Dauser} \& {Gallo}]{Vasudevan16}
{Vasudevan} R.~V., {Fabian} A.~C., {Reynolds} C.~S., {Aird} J., {Dauser} T.,
  {Gallo} L.~C., 2016, \mnras, 458, 2012

\bibitem[{Verner} et~al.(1996){Verner}, {Ferland}, {Korista} \&
  {Yakovlev}]{Verner96}
{Verner} D.~A., {Ferland} G.~J., {Korista} K.~T., {Yakovlev} D.~G., 1996, \apj,
  465, 487

\bibitem[{Walton} et~al.(2018){Walton}, {Brightman}, {Risaliti}
  et~al.]{Walton18}
{Walton} D.~J., {Brightman} M., {Risaliti} G., et~al., 2018, \mnras, 473, 4377

\bibitem[{Walton} et~al.(2013){Walton}, {Nardini}, {Fabian}, {Gallo} \&
  {Reis}]{Walton13spin}
{Walton} D.~J., {Nardini} E., {Fabian} A.~C., {Gallo} L.~C., {Reis} R.~C.,
  2013, \mnras, 428, 2901

\bibitem[{Walton} et~al.(2019){Walton}, {Nardini}, {Gallo} et~al.]{Walton19ufo}
{Walton} D.~J., {Nardini} E., {Gallo} L.~C., et~al., 2019, \mnras, 484, 2544

\bibitem[{Walton} et~al.(2012){Walton}, {Reis}, {Cackett}, {Fabian} \&
  {Miller}]{Walton12xrbAGN}
{Walton} D.~J., {Reis} R.~C., {Cackett} E.~M., {Fabian} A.~C., {Miller} J.~M.,
  2012, \mnras, 422, 2510

\bibitem[{Walton} et~al.(2015){Walton}, {Reynolds}, {Miller}, {Reis}, {Stern}
  \& {Harrison}]{Walton15lqso}
{Walton} D.~J., {Reynolds} M.~T., {Miller} J.~M., {Reis} R.~C., {Stern} D.,
  {Harrison} F.~A., 2015, \apj, 805, 161

\bibitem[{Walton} et~al.(2014){Walton}, {Risaliti}, {Harrison}
  et~al.]{Walton14}
{Walton} D.~J., {Risaliti} G., {Harrison} F.~A., et~al., 2014, \apj, 788, 76

\bibitem[{Walton} et~al.(2016){Walton}, {Tomsick}, {Madsen}
  et~al.]{Walton16cyg}
{Walton} D.~J., {Tomsick} J.~A., {Madsen} K.~K., et~al., 2016, \apj, 826, 87

\bibitem[{Wilkins} \& {Fabian}(2012)]{Wilkins12}
{Wilkins} D.~R., {Fabian} A.~C., 2012, \mnras, 424, 1284

\bibitem[{Wilms} et~al.(2000){Wilms}, {Allen} \& {McCray}]{tbabs}
{Wilms} J., {Allen} A., {McCray} R., 2000, \apj, 542, 914

\bibitem[{Wilson} \& {Colbert}(1995)]{Wilson95}
{Wilson} A.~S., {Colbert} E.~J.~M., 1995, \apj, 438, 62

\bibitem[{Winkler} et~al.(2003){Winkler}, {Courvoisier}, {Di Cocco}
  et~al.]{INTEGRAL}
{Winkler} C., {Courvoisier} T.~J.~L., {Di Cocco} G., et~al., 2003, \aap, 411,
  L1

\bibitem[{Winter} et~al.(2009){Winter}, {Mushotzky}, {Reynolds} \&
  {Tueller}]{Winter09}
{Winter} L.~M., {Mushotzky} R.~F., {Reynolds} C.~S., {Tueller} J., 2009, \apj,
  690, 1322

\bibitem[{Zdziarski} et~al.(1996){Zdziarski}, {Johnson} \&
  {Magdziarz}]{nthcomp1}
{Zdziarski} A.~A., {Johnson} W.~N., {Magdziarz} P., 1996, \mnras, 283, 193

\bibitem[{Zhang} et~al.(2019){Zhang}, {Dov{\v{c}}iak} \& {Bursa}]{Zhang19}
{Zhang} W., {Dov{\v{c}}iak} M., {Bursa} M., 2019, \apj, 875, 2, 148

\bibitem[{Zoghbi} et~al.(2012){Zoghbi}, {Fabian}, {Reynolds} \&
  {Cackett}]{Zoghbi12}
{Zoghbi} A., {Fabian} A.~C., {Reynolds} C.~S., {Cackett} E.~M., 2012, \mnras,
  422, 129

\bibitem[{Zycki} et~al.(1999){Zycki}, {Done} \& {Smith}]{nthcomp2}
{Zycki} P.~T., {Done} C., {Smith} D.~A., 1999, \mnras, 309, 561

\end{thebibliography}

\label{lastpage}

\end{document}